\documentclass[onecolumn, notitlepage,superscriptaddress,amsmath,amssymb,longbibliography,aps,floatfix]{revtex4-1}
\usepackage{graphicx,bm}
\usepackage{bm}
\usepackage{color}
\usepackage[breaklinks,colorlinks = true,linkcolor = blue,urlcolor=blue,citecolor=blue]{hyperref}
\usepackage{cancel}
\usepackage[normalem]{ulem}
\usepackage{amsfonts}
\usepackage{amsmath}
\usepackage{hyperref}
\usepackage{romannum}
\usepackage[dvipsnames]{xcolor}
\graphicspath{{./}{fig/}}
\def \Rey  {\mbox{Re}}
\def \Rel  {\Rey_\lambda}
\def \Deb  {\mbox{De}}	
\def \urms {u_{\rm rms}}
\def \dV   {\delta V}

\def \sp   {S_{\rm p}}
\def \taup {\tau_{\rm p}}
\def \tauL {\tau_{\rm L}}
\def \rp   {\rm p}
\def \rq   {\rm q}
\def \Stq  {\tilde{S}_{\rq}}
\def \Stp  {\tilde{S}_{\rp}}
\def \br   {{\bm r}}
\def \bx   {{\bm x}}
\def \bh   {{\bm h}}
\def \bxp  {{\bm x}'}
\def \bX   {{\bm X}}

\def \bu   {{\bm u}}
\def \dbu  {\delta \bu}
\def \diffu {u(x+r) - u(x)}
\def \avgT {\frac{T(x+r) + T(x)}{2}}

\def \bmf   {{\bm f}}

\def \rD   {{\rm D}}
\def \rF   {{\rm F}}
\def \rd   {{\rm d}}

\def \diss {\varepsilon}
\def \dissf {\varepsilon_{\rm f}}
\def \dissi {\varepsilon_{\rm I}}
\def \dissfr {\varepsilon_{{\rm f},r}}
\def \disst {\varepsilon_{\rm t}}

\def \dissp {\varepsilon_{\rp}}
\def \disspr {\varepsilon_{{\rm p},r}}
\def \dissr {\varepsilon_r}
\def \disstr {\varepsilon_{{\rm t},r}}
\def \zetap {\zeta_{\rm p}}
\def \xip {\xi_{\rm p}}

\def \intS {\int_{\rm{S(r)}} dS}
\def \intV {\int_{\Omega(r)} d\Omega \,}

\def \fflux {\Phi_{\rm f}}
\def \pflux {\Phi_{\rm p}}
\def \tflux {\Phi_{\rm t}}

\def \hot {\text{h.o.t.}}

\def \calp  {\mathcal{P}}
\def \Rij {R_{ij}}
\def \Rii {R_{ii}}
\def \Rik {R_{ik}}
\def \Rkj {R_{kj}}

\def \pt {\frac{\partial}{\partial t}}
\newcommand \pX[1] {\frac{\partial}{\partial X_{#1}}}
\newcommand \pr[1] {\frac{\partial}{\partial r_{#1}}}
\newcommand \pXX[1] {\frac{\partial^2}{\partial X^2_{#1}}}
\newcommand \prr[1] {\frac{\partial^2}{\partial r^2_{#1}}}

\newcommand \pxi[2] {\frac{\partial #2}{\partial x_{#1}}}
\newcommand \psxi[2] {\frac{\partial^2 #2}{\partial x_{#1}^2}}

\newcommand \lrp[1] {\left( #1 \right)}
\newcommand \lrs[1] {\left[ #1 \right]}
\newcommand \lrv[1] {\left\lvert #1 \right\rvert}

\newcommand \bra[1] {\left\langle #1 \right\rangle}

\newcommand \delu[1]   {\delta u_{#1}}
\newcommand \delus[1]   {\lrp{\delta u_{#1}}^2}

\newcommand \Stilde[1]  {\tilde{S}_{#1}}
\newcommand \Sf[1]   {S_{#1}}

\newcommand \ints[1] {\int_{\rm{S(r)}} dS \, n_{#1}}
\newcommand \disstlm[1] {\left\langle \varepsilon_{{\rm t},{\rm r}}^{#1/3} \right\rangle }
\newcommand \Ht[2]  {\tilde{H}_{#1,#2}}

\newcommand \mult[1]   {\alpha_{#1} \lrp{\bX,r,r'}}

\newcommand{\oist}{Complex Fluids and Flows Unit, Okinawa Institute of Science and Technology Graduate University, Okinawa 904-0495, Japan}

\begin{document}

\title{Extending Kolmogorov Theory to Polymeric Turbulence} 
\author{Alessandro Chiarini}
\altaffiliation{Present address: Dipartimento di Scienze e Tecnologie Aerospaziali, Politecnico di Milano, via La Masa 34, 20156 Milano, Italy}
\email{alessandro.chiarini@polimi.it}
\affiliation{\oist}
\author{Rahul K. Singh}
\email{rksphys@gmail.com}
\affiliation{\oist}
\author{Marco E. Rosti}
\email{marco.rosti@oist.jp}
\affiliation{\oist}

\begin{abstract}
The addition of polymers fundamentally alters the dynamics of turbulent flows in a way that defies Kolmogorov predictions. However, we now present a formalism that reconciles our understanding of polymeric turbulence with the classical Kolmogorov phenomenology. This is achieved by relying on an appropriate form of the K\'{a}rm\'{a}n-Howarth-Monin-Hill relation, which motivates the definition of extended velocity increments and the associated structure functions, by accounting for the influence of the polymers on the flow. We show, both analytically and numerically, that the $\rp{th}$-order extended structure functions exhibit a power-law behaviour in the elasto-inertial range of scales, with exponents deviating from the analytically predicted value of $\rp/3$. These deviations are readily accounted for by considering local averages of the total dissipation, rather than global averages, in analogy with the refined similarity hypotheses of Kolmogorov for classical Newtonian turbulence. We also demonstrate the scale-invariance of multiplier statistics of extended velocity increments, whose distributions collapse well for a wide range of scales.
\end{abstract}

\maketitle

\section{Introduction}

The celebrated works of Kolmogorov form the basis of our understanding of homogeneous, isotropic turbulence which has been a long standing problem in classical physics~\citep{Lvov1996,Ecke2005}. In his 1941 work (K41), Kolmogorov showed that, in a stationary state of fully developed turbulence, the skewness $\Sf{3} \equiv \bra{ (\dbu \cdot \br/r )^3}$ of the distribution of the velocity increments $\dbu = \bu( \bx + \br) - \bu(\bx)$ is related to the average rate of the energy dissipation $\bra{\diss}$ as $\Sf{3} = (-4/5) \bra{\diss} r$, where $r = \lrv{\br}$~\citep{K41c,Frisch96,Pope2000}. This universal relation holds in the so-called inertial range of scales $\eta \ll r \ll L$, where $L$ is the (large) scale of energy injection and $\eta$ is the (small) dissipative Kolmogorov scale. This exact result is central to any theory of turbulence, and emerges as a direct consequence of the K\'{a}rm\'{a}n--Howarth relation, which states that energy is, on an average, transferred from large to small scales at a constant rate, and equals the rate of energy dissipation~\citep{karman-howarth-1938}. A phenomenological extension of this relation via dimensional arguments suggests that a general $\rp$th-order structure function must scale as $\sp \equiv \bra{\lrp{\dbu \cdot \br/r}^{\rp}} \sim \bra{\diss}^{\rp/3} r^{\rp/3}$~\citep{Frisch96,Pope2000}. The universality of this result was however questioned by Landau, who argued that the random nature of the energy transfer from large to small scales can not be disregarded~\citep{landau-lifschitz-1959,Frisch96}. This led Kolmogorov, following a suggestion by Oboukhov, to refine the similarity hypotheses (KO62) and account for the intermittent nature of turbulence by relaxing the global averaging of dissipation $\bra{\diss}$ in favour of its locally averaged value over a scale $r$, $\bra{\dissr}$, so that now $\sp \sim \bra{ \diss_r^{\rp/3} } r^{\rp/3}$~\citep{K62}. In doing so, Kolmogorov hinted that the intermittency of the velocity fluctuations originates from an intermittent energy flux, which culminates in an intermittent energy dissipation. Although a first principles' derivation of this revised relation remains an open problem, numerical and experimental studies have underlined its validity and role in understanding small-scale intermittency in Newtonian turbulence~\citep{stolovitzky-etal-1992,praskovsky-1992,Antonia1995,Nelkin1997,Iyer2015,Bodenschatz2019,Yeung2024}. 

The Kolmogorov phenomenology describes the nature of turbulence in flows of Newtonian fluids. In general, however, turbulent flows can be multiphase and transport objects with various internal and external degrees of freedom. Such added phases, by virtue of their interaction with the carrier flow, modify the nature of turbulence in a non-trivial manner~\citep{Voth2017,Benzi2018,Brandt2022}. A particularly interesting situation arises when polymers are added to a carrier flow. A small concentration of polymers in a turbulent flow, indeed, modifies the way energy is distributed and transferred across scales~\citep{Casciola2007,zhang-etal-2021,MER2023}, and results in a variety of intriguing, emergent phenomena, that range from a chaotic state of `elastic turbulence' in otherwise laminar flows~\citep{Steinberg2021,Singh2024} to that of drag reduction in turbulent flows~\citep{Godfrey2008}. The nature of these polymeric flows is characterised by two dimensionless numbers, the Reynolds number $\Rey$ and the Deborah number $\Deb$. The Reynolds number $\Rey \equiv L \urms/\nu$ is a measure of the rough, chaotic and unpredictable nature of the flow, where $\urms$ is the root-mean-square of the velocity fluctuations, and $\nu$ the kinematic viscosity of the fluid. A flow is commonly said to be turbulent when $\Rey$ is large enough, so that the Kolmogorov theory holds in the asymptotic limit of $\Rey \to \infty$. The Deborah number $\Deb \equiv \taup/\tauL$ quantifies the elasticity of the polymers via a typical polymer relaxation time $\taup$ relative to the largest time-scale of the flow $\tauL \equiv L/\urms$. A large $\Deb$ implies that polymers stretch more and take longer to relax back to their equilibrium configuration. Polymeric flows at large $\Rey$ and unit $\Deb$ have recently been shown, both experimentally and numerically~\cite{zhang-etal-2021,MER2023}, to exhibit a novel, self-similar scaling regime, where $\Sf{2} \sim r^{\xi_2}$ with $\xi_2 \approx 1.3$, which departs significantly from the Kolmogorov exponent of $\xi_2 = 2/3$. This can be seen in Fig.~\ref{fig:Fig2}(a) where the intermediate $2/3$-rd scaling range in Newtonian turbulence is modified to a $1.3$ scaling range in polymeric turbulence.

\begin{figure}
	\centering	    
    \includegraphics[width=\columnwidth]{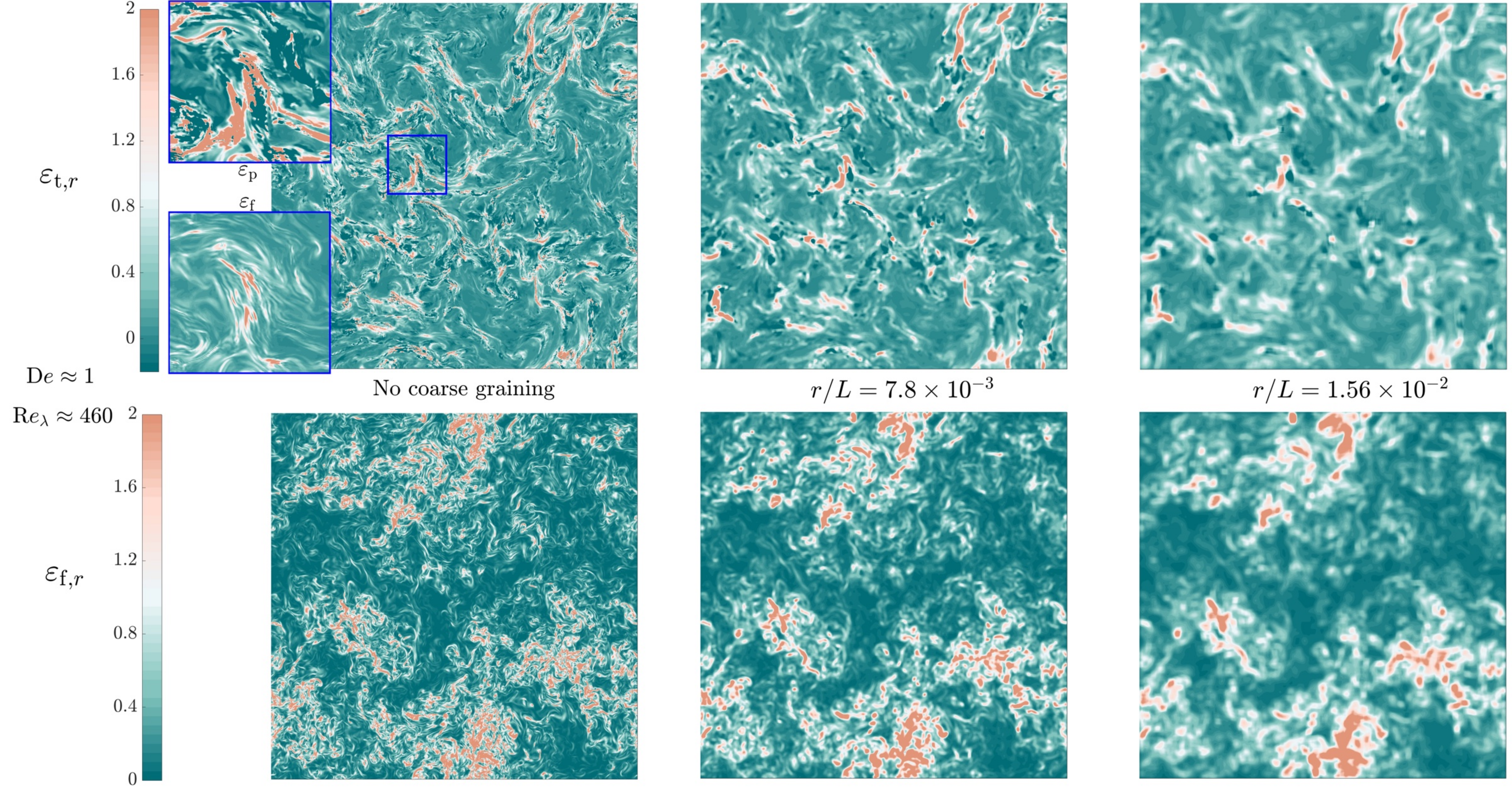}	
	\caption{{\bf Snapshots from simulations}. {\bf Top:} Total dissipation $\disstr$ for our turbulent polymeric flows with $\Deb \approx 1$, at different levels of local coarse graining $r$. % where $\Delta x = 2\pi/1024$ is the smallest grid separation.
The leftmost panel shows zoomed in views of polymer $\disspr$ and fluid $\dissfr$ dissipation rates, within the section marked by the blue square. We have $\disstr = \disspr + \dissfr$. {\bf Bottom:} Locally averaged fluid dissipation $\dissfr$ from a simulation of a purely Newtonian fluid at the same $\Rel \approx 460$. Here, $\disstr = \dissfr$.}
	\label{fig:Fig1}
\end{figure}

It is clear that, the addition of polymers drives turbulence statistics far away from the classical Kolmogorov predictions. However, we show in this work that turbulence statistics for polymeric flows can be cast within a K41-like phenomenology, with suitable modifications following directly from the governing equations of motion. We also show that the refined KO62 phenomenology accounts for appropriate corrections to the deviations from a K41-like behaviour, similar to classical Newtonian turbulence. In what follows, we begin by setting up the analytical framework, and substantiate it later with data from numerical simulations.

\section{The Kármán-Howarth-Monin-Hill Relation} 
\label{sec:derivation}

We begin our discussion by building the theoretical background. The K\'{a}rm\'{a}n-Howarth-Monin-Hill (KHMH) relation follows directly from the Navier--Stokes equations, and describes how energy is transferred across various scales of motion in a turbulent flow~\citep{Frisch96,Pope2000}. Several variants of the KHMH equation, also known as generalised Kolmogorov equations, have been used to understand the scale-space energy transfers in anisotropic and inhomogeneous turbulent flows~\citep{danaila-etal-2001,marati-casciola-piva-2004,cimarelli-deangelis-casciola-2013,cimarelli-etal-2016,yasuda-vassilicos-2018,alvesportela-papadakis-vassilicos-2020,gatti-etal-2020,chiarini-etal-2021,yao-etal-2024b}. We now obtain an equivalent KHMH relation (see also~\citep{deangelis-etal-2005}) for turbulent polymeric flows, starting from the governing equations of motion. As we shall see later, this naturally motivates an extension of the Kolmogorov phenomenology to polymeric turbulence. 

The dynamics of polymeric flows is governed by the modified Navier-Stokes equations, that incorporate an additional polymer stress term to model how the polymers react to the carrier flow, coupled with an evolution equation for polymer stresses:
\begin{align}
  \frac{\partial u_{i}}{\partial t} &+ u_{j} \pxi{j}{u_i}  =   - \frac{1}{\rho}\pxi{i}{p} + \nu \psxi{j}{u_i} + 
 \frac{1}{\rho} \pxi{j}{T_{ij}} + f_i \ ; 	\qquad  		\pxi{j}{u_j} = 0	\ ,																	  \label{eq:MNS} \\
  \frac{\partial \Rij}{\partial t} &+ u_k \pxi{k}{\Rij}  =  \pxi{k}{u_i} \Rkj+ \Rik \pxi{j}{u_k} -
  \frac{1}{\tau_p} \left( \calp \Rij - \delta_{ij} \right). 											\label{eq:Tab} 
\end{align}  
Here, $\bu$ is the incompressible fluid velocity, $p$ is the pressure, $\nu$ is the kinematic fluid viscosity, $\rho = 1$ is the fluid density, and ${\bm f}$ is the forcing that sustains turbulence. The polymer-stress forces on the carrier fluid are given by the derivative of the symmetric, polymeric stress tensor $T_{ji} \equiv T_{ij}$, which is related to the polymer conformation tensor $R_{ij}$ as $R_{ij} = T_{ij} \tau_p /\mu_p + \delta_{ij}$, where $\tau_p$ is the polymer relaxation time and $\mu_p$ the polymeric viscosity. $\mathcal{P}$ is the Peterlin function, so that $\mathcal{P}=1$ corresponds to the purely elastic Oldroyd-B model and $\mathcal{P} = ( \ell_{\max}^2 -3 )/( \ell_{\max}^2 - R_{ii} )$ to the FENE-P model, where $\ell_{\max}$ is the maximum polymer extensibility and $\Rii$ is the instantaneous squared end-to-end length of the polymers. Eqns.~\eqref{eq:MNS} and \eqref{eq:Tab} are supplemented with periodic boundary conditions for  the following analysis, and we assume a statistically stationary state of (forced) fully developed polymeric turbulence. %We present our results for $\Rel \equiv \urms \lambda / \nu \approx 460$ ($\lambda$ is the Taylor microscale) and $\Deb \approx 1$.

We start by writing Eqn.~\eqref{eq:MNS} at two distinct points $\bx = \bX + \br/2$, $\bxp = \bX - \br/2$, separated by $\br$ and centred at $\bX$. The difference of the two equations yields an evolution equation for the velocity increment $\delu{i} \lrp{\bX,\br, t} = u_i (\bx',t) - u_i (\bx,t)$ over a separation $\br$ and about the point $\bX$, where $i = 1,2,3$ runs over the three Cartesian directions. The energy equation then results upon a multiplication by $\delu{i}$:
\begin{align}
  \pt \delu{i}^2  = & - \pX{j} \lrp{u_j^* \delu{i}^2} - 2 \pX{j} \lrp{\delta p \delu{j} } + \pX{j} \lrp{\delta T_{ij} \delu{i}}  + \frac{\nu}{2} \pXX{j} \delu{i}^2 \nonumber \\ 
  & - \pr{j} \lrp{\delu{j} \delu{i}^2 } + 4 \pr{j} \lrp{\delu{i} T_{ij}^*} + 2 \nu \prr{j} \delu{i}^2  - 4  \nu \left( \frac{\partial u_i}{\partial x_j}\frac{\partial u_i}{\partial x_j} \right)^*   - 4 \lrp{ T_{ij} \frac{\partial u_i}{\partial x_j} }^*  + 2 \delta f_i \delu{i},
  \label{eq:duidui}
\end{align}
where repeated indices are summed, and $\cdot^*$ denotes the average of a given quantity at the points $\bm{x}$ and $\bm{x}'$, e.g. $u_j^* = \left( u_j(\bm{x}) + u_j(\bm{x}') \right)/2$. Eqn.~\eqref{eq:duidui} captures the instantaneous rate of change of energy at any point, both in space and scales,  resulting from the transfers of energy due to fluid-fluid and fluid-polymer interactions, as well as those due to dissipation. The KHMH equation is then obtained by averaging $\lrp{\bra{\cdot}}$ over the ensemble of realisations and the three homogeneous directions, i.e. the total volume. Periodicity in space implies that the spatial $\partial/\partial X_j$ derivatives result in a zero contribution, and that only the scale derivatives $\partial/\partial r_j$ survive. Moreover, assumption of statistical stationarity means $\partial \bra{\cdot}/\partial t = 0$. The KHMH equation for homogeneous, polymeric turbulence thus reads
\begin{align}
   \pr{j} \bra{\delu{j} \delu{i}^2} - 4 \pr{j} \bra{ \delu{i} T_{ij}^* } -2 \nu \prr{j} \bra{ \delu{i}^2 } =  - 4 \bra{\dissf} - 4 \bra{\dissp} + 2 \bra{ \delta f_i \delu{i} }.
 \label{eq:KHMH-hom}
\end{align}
Here, $\bra{\dissp} \equiv \bra{T_{ij} \partial u_i/ \partial x_j  } = \bra{T_{jj}}/2\taup = \lrp{\mu_p/2\taup^2} \bra{R_{jj} - 3}$, where $T_{jj}(\bx,t)$ is the local, positive-definite polymeric dissipation~\citep{deangelis-etal-2005}. However, the local $\dissp$ can be both positive or negative unlike the positive definite fluid pseudo-dissipation $\dissf = \nu (\partial u_{i}/\partial x_{j})^2$. This means that polymers can locally either force or dampen the velocity fluctuations, while globally removing energy from the flow. We now integrate Eqn.~\eqref{eq:KHMH-hom} in a spherical volume of radius $r$, thus converting it into a scalar relation for the net outward flux of energy from a scale $r$ as
\begin{align}
  \ints{j} \bra{ \delu{j} \delus{i} } +  \ints{j} \bra{ - 4 \delu{i} T_{ij}^* }  &= \nonumber \\
  2 \nu \ints{j} \pr{j} \bra{ \delus{i} }  &+  2 \intV \bra{\delta f_i \delu{i} }   - 4 \intV \bra{ \dissf } - 4 \intV \bra{ \dissp } .
\end{align}
The volume integrals $\intV$ have been switched to integrals over the bounding surface $\intS$ using the Gauss divergence theorem, with the unit normal denoted by $\bm{n}$. Dividing by the surface area $S(r) = 4 \pi r^2$ to obtain local averages, we finally have
\begin{align}
\underbrace{ \frac{1}{4 \pi r^2}  \ints{j} \bra{ \delu{j} \delus{i} }}_{\fflux(r)}  +
\underbrace{ \frac{1}{4 \pi r^2} \ints{j} \bra{ - 4 \delu{i} T_{ij}^* } }_{\pflux(r)} = &
\underbrace{  \frac{2\nu}{4 \pi r^2} \ints{j} \pr{j} \bra{ \delus{i} }}_{\text{D}(r)} +
\underbrace{ \frac{2}{4 \pi r^2} \intV \bra{\delta f_i \delu{i} } }_{{\rm F}(r)} 	\nonumber \\
&\underbrace{- \frac{4}{3} \bra{ \diss_f } r - \frac{4}{3} \bra{ \diss_p } r}_{-\frac{4}{3} \bra{\diss_t} r},
\label{eq:KHMH}
\end{align}
where $\diss_t = \diss_f + \diss_p$ is the total local dissipation. This is the generalisation of the K\'{a}rm\'{a}n-Howarth equation for homogeneous turbulent, polymeric flows. 

We can now make even further simplifying assumptions by focusing our attention on an intermediate range of scales. We assume that this range is sufficiently far from the limited band at large $r$ where the external forcing $\bmf$ injects energy into the flow, as well as from the smallest scales where the viscous effects are relevant. Consequently, the forcing $\text{F}(r)$ and viscous $\text{D}(r)$ contributions can be neglected in this `elasto-inertial' range, so that Eqn.~\eqref{eq:KHMH} reduces to
\begin{align}
  \tflux(r) \equiv \fflux(r) + \pflux(r) = - \frac{4}{3} \bra{\diss_t} r,
  \label{eq:KHMH_ir}
\end{align}
where $\fflux$, $\pflux$ are the fluid non-linear (inertial) and the non-Newtonian (elastic) fluxes, respectively, while $\disst$ is the total local dissipation. Eqn.~\eqref{eq:KHMH_ir} reveals the clear distinction between Newtonian and polymeric turbulence~\cite{MER2023}. In turbulent flows of Newtonian fluids, dissipation $\bra{\dissf}$ results only due to the fluid viscosity, and the flux of energy through scales is only due to the fluid non-linearity $\fflux$. Turbulent polymeric flows, however, present an additional dissipation $\bra{\dissp}$, and an additional flux contribution $\pflux$ due to the polymeric microstructure. Eqn.~\eqref{eq:KHMH_ir} further hints that the rate of \textit{total} energy transfer $\rd \tflux/\rd r$ is invariant in the elasto-inertial range of scales, and equals the \textit{total} rate of energy dissipation $\bra{ \disst } = \bra{ \dissf} + \bra{ \dissp }$.

A further assumption of local statistical isotropy yields the equivalent of the $4/5$-th law of Kolmogorov in polymeric turbulence. Indeed, we can write the left side of Eqn.~\eqref{eq:KHMH} for isotropic statistics as
\begin{align}
  \frac{1}{4 \pi r^2} \ints{j} \bra{ \delu{j} \delus{i} }   = \bra{\delu{\parallel} \delus{i} } \quad \text{and} \quad
   \frac{1}{4 \pi r^2 } \ints{j} \bra{-4 \delu{i} T_{ij}^* }  = - 4 \bra{\delu{i} T_{i\parallel}^* },
\end{align}
where the subscript $\cdot_\parallel$ denotes the direction parallel to $\bm{r}$. Eqn.~\eqref{eq:KHMH_ir}, therefore, can be recast as
\begin{align}
  \bra{\delta u_\parallel (\delta u_i)^2} - 4 \bra{\delta u_i T_{i\parallel}^* } = - \frac{4}{3} \bra{\diss_t} r.
  \label{eq:43law}
\end{align}
This relation can be further simplified by using the incompressibility condition $r \partial \bra{ \delta u_\parallel^3 }/\partial r + \bra{ \delta u_\parallel^3} - 6 \bra{\delta u_\parallel \delta u_\perp^2} = 0$ to obtain~\citep{hill-1997,hill-2002}
\begin{align}
  \frac{r}{3} \pr{} \bra{\delu{\parallel}^3 } +  \frac{4}{3} \bra{\delu{\parallel}^3} -  4 \bra{ \delu{i} T_{i\parallel}^* } =  - \frac{4}{3} \bra{\disst} r.
\end{align}
Multiplying the above equation by $3r^3$ and integrating, one eventually obtains:
\begin{align}
  \bra{\delu{\parallel}^3} - \frac{12}{r^4} \int_0^r \text{d}\ell \ \ell^3 \bra{\delu{i} T_{i\parallel}^* }  = - \frac{4}{5} \bra{\disst} r.
  \label{eq:45law}
\end{align}
Eqn.~\eqref{eq:45law} is the polymeric equivalent of Kolmogorov's celebrated $4/5$-th law in Newtonian turbulence. This relation indicates a statistical equivalence of two ergodic processes: the averaged total dissipation is directly related to the sum of the third moment of the longitudinal velocity increment and a fluid-polymeric coupling term.  

We are now in the position to extend the Kolmogorov phenomenology to polymeric turbulence. To this end, we introduce the extended structure function (ESF) of order $\rp$ as $\Stp \equiv \bra{ \widetilde{\dV}_\parallel^{\rp}}$, where $\widetilde{\dV}_\parallel$ is the extended longitudinal velocity increment whose definition arises naturally from Eqn.~\eqref{eq:43law}:
\begin{align}
  \widetilde{\dV}_\parallel = \lrp{ \delu{\parallel} \delus{i} - 4 \delu{i} T_{i\parallel}^*} ^{1/3},
  \label{eq:dvt}
\end{align}
where the cube root is defined as the real signed root of the expression within the parenthesis. Clearly, from Eqn.~\eqref{eq:45law} we have that $\widetilde{\delta V}_\parallel = \delta u_\parallel$ in the absence of any dissolved polymers, so that $\Stp$ reduces to the classical $\Sf{\rp}$. A qualitatively similar observable, though with very different contributions, was also identified for Newtonian, homogeneous shear turbulence~\citep{casciola-etal-2003}. 

The extended observable $\widetilde{\dV}_\parallel$ captures both the inertial (Newtonian) and the elastic (non-Newtonian) contributions to the total flux. Notably, unlike the conventional velocity increments, $\widetilde{\dV}_\parallel$ is not simply a difference, as the non-Newtonian contribution also comprises an average between two points $T_{ij}^*$. Nonetheless, in the spirit of K41, Eqn.~\eqref{eq:43law} inspires us to use dimensional arguments and expect for any $\rp$th-order ESF $\Stp$ a power-law dependence on $r$ as 
\begin{align}
 \Stp  \sim \bra{ \disst }^{\rp/3} r^{\rp/3}.
 \label{eq:PK41}
\end{align}
Such a relation is expected to hold exactly only for $\rp=3$. The corrections to such a relation, in analogy with KO62, may be accounted for by considering the locally scale-averaged dissipation $\disstr \lrp{\bx}$  $\equiv$  $(4\pi r^3/3)^{-1} \int_{\lrv{\bh}  \leq r} \disst (\bx + \bh) d\bh$, rather than the global average $\bra{\disst}$, to obtain
\begin{align}
  \Stp \sim \bra{ \disstr^{\rp/3} } r^{\rp/3}.				\label{eq:PK61}
\end{align} 

\section{Numerical Evidence}

\begin{figure}
	\centering
	\includegraphics[width=\textwidth]{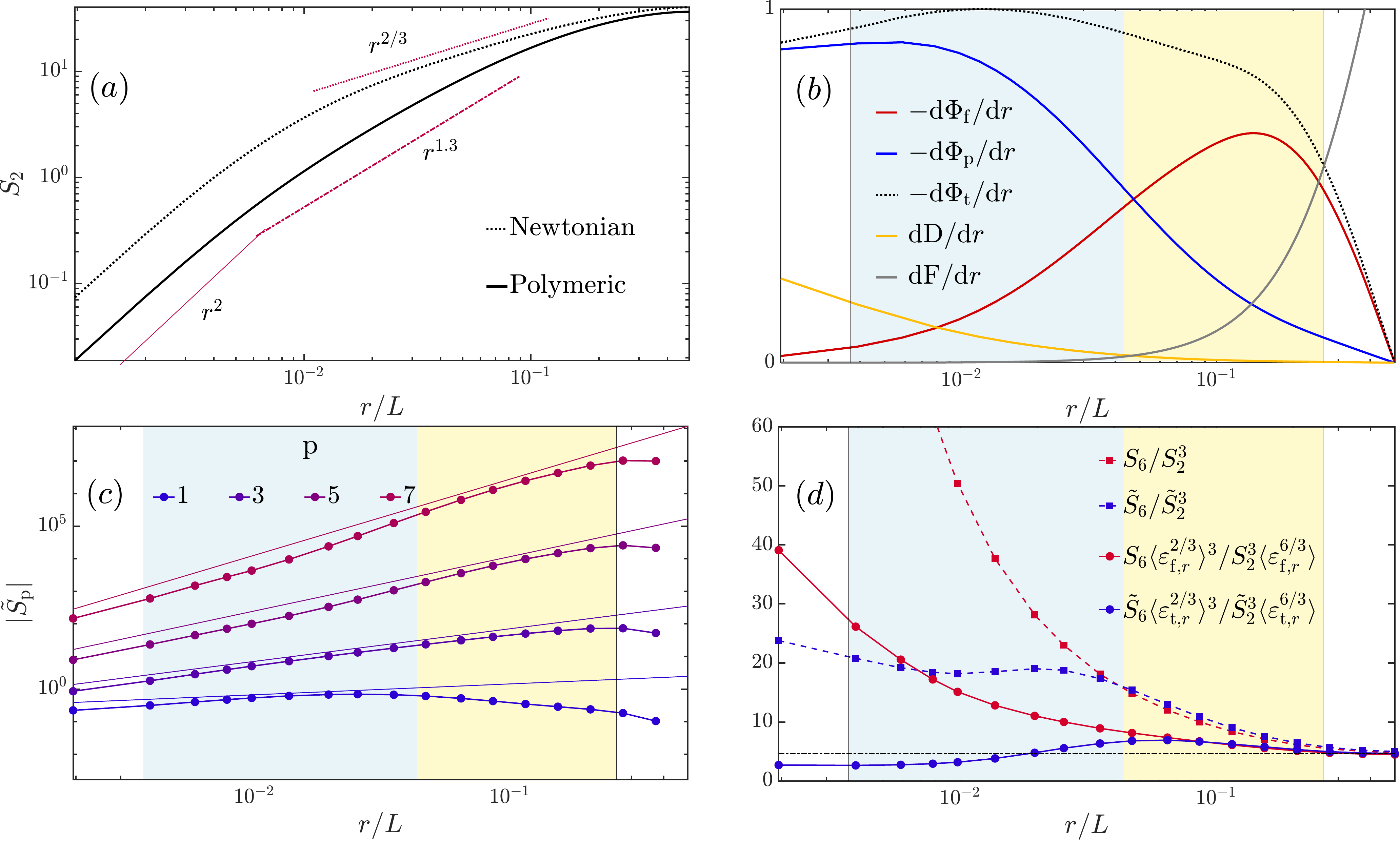}	
	\caption{\textbf{Structure Functions and Fluxes.} \textbf{(a)} The fluid velocity structure functions $\Sf{2}$ in polymeric turbulence (solid black curve) shows a scaling of the form $\Sf{2} \sim r^{1.3}$~\cite{zhang-etal-2021,MER2023}, in a marked deviation from that for Newtonian turbulence (dotted black curve), for which $\Sf{2} \sim r^{2/3}$. \textbf{(b)} This novel $r^{1.3}$ scaling holds in an intermediate range where fluid and polymeric contributions sum to a constant total energy transfer rate $\rd \tflux /\rd r = \rd \fflux /\rd r + \rd \pflux /\rd r$. We show in blue (yellow) the range of scales dominated by $\pflux$ ($\fflux)$. The dissipative $\rD$ and forcing $\rF$ contributions remain small in this entire range. \textbf{(c)} The Extended Structure Functions $\Stp$ in polymeric turbulence exhibit a close to Kolmogorov behaviour $\Stp\sim r^{\rp/3}$. The solid lines without symbols show the exact $\rp/3$ scaling. \textbf{(d)} The Refined Kolmogorov similarity hypothesis hold in polymeric turbulence. This is apparent from the flat compensated curve $\Stilde{6} \bra{\disstr^{2/3}}^3/\Stilde{2}^3 \bra{\disstr^{6/3}}$ shown in blue circle markers. In comparison, the Newtonian compensated (solid red, circle markers) and the uncompensated (dashed red, square markers) ratios, as well as the uncompensated extended ratio (dashed blue curve, square markers), are strongly dependent on the scale $r$.}
	\label{fig:Fig2}
\end{figure}
Having set up the analytical arguments, we now substantiate our claims using numerical evidence, and show that polymeric turbulence can indeed be reconciled with a Kolmogorov-like phenomenology. To this end, we use data from direct numerical simulations of homogeneous, isotropic, polymeric turbulence at a microscale Reynolds number of $\Rel = \urms \lambda/\nu \approx 460$, with $\lambda$ being the Taylor length scale, the details of which can be found in Ref.~\cite{RKS2023}. The polymeric phase is modelled as an Oldroyd-B fluid, with a relaxation time-scale giving $\Deb \approx 1$, where the non-Kolmogorovean nature of velocity statistics is manifested over a maximal range of scales (also shown in  Fig.~\ref{fig:Fig2}(a))~\citep{MER2023,RKS2023}. We show the scale averaged total $\disstr=\dissfr+\disspr$ dissipation from representative two-dimensional slices of our simulations in Fig.~\ref{fig:Fig1}, at different levels of scale averaging. We discuss the influence of the polymer elasticity  and the model independence of our results in the Supplementary Materials.

\subsection{Extended Structure Functions}

We begin by emphasising again the deviation of statistics in polymeric turbulence from the Kolmogorov predictions. This is evident in Fig.~\ref{fig:Fig2}(a) where the second-order structure function $\Sf{2}$ in an intermediate range of scales clearly shows a non-Kolmogorovean scaling $\Sf{2} \sim r^{1.3}$ for turbulent polymeric flows~\citep{zhang-etal-2021,MER2023}. Turbulent Newtonian flows at this Reynolds numbers exhibit instead the classical Kolmogorov scaling $\Sf{2} \sim r^{2/3}$.

The KHMH relation~\eqref{eq:KHMH_ir} implies that a constant rate of energy transfer $\rd \tflux/\rd r $ in the elasto-inertial range results in 
a Kolmogorov-like scaling given by Eqn.~\eqref{eq:PK41} for our ESFs $\Stp$. Such a prediction holds very well, as seen from the plots of (the absolute of) odd order $\Stp$ in Fig.~\ref{fig:Fig2}(c), and especially for the exact case of $\rp = 3$. We show that this relation holds even more closely in the Supplementary Material, by plotting $\lrv{\Stilde{3}}/\bra{\disst} r$ which stays close to $4/3$ for a wide range of scales. %The straight lines show the expected $\rp/3$ scaling to guide the eye. 
This is the first confirmation of our hypotheses: K41 scaling holds in polymeric turbulence for an extended observable $\widetilde{\dV}_\parallel$. The underlying assumptions of the forcing ${\rm F}(r)$ and the viscous ${\rm D}(r)$ contributions being negligible in the elasto-inertial range are justified from the corresponding curves in panel (b). Forcing is non-negligible at the large scales while the viscous term is appreciable at only the smallest scales. Furthermore, in the intermediate range, a dominant fluid non linearity $\text{d} \fflux/\text{d} r$ marks the inertial range at large $r$, while a dominant polymeric contribution $\text{d} \pflux/\text{d} r$ defines the elastic range at relatively small $r$. However, in this entire elasto-inertial range, the net rate of energy transfer $\text{d} \tflux /\text{d} r = -4/3 \bra{\disst}$ is approximately constant, shown as the dotted black curve in Fig.~\ref{fig:Fig2}(b). This explains the observed $\Stp \sim r^{\rp/3}$ scaling for intermediate $r$ in Fig.~\ref{fig:Fig2}(c). Unexpectedly, we also have that the same $\rp/3$ scaling holds even for $r \to 0$, where dissipation clearly dominates and Eqn.~\eqref{eq:KHMH_ir} no longer holds. Such a behaviour at small $r$ is, in fact, a consequence of the analytic nature of the velocity and the polymeric fields. We discuss in detail in the Supplementary Materials how this analytic nature implies a dominant non-Newtonian contribution, yielding a $\rp/3$ power-law behaviour. The ESFs therefore exhibit two distinct regimes of the $\rp/3$ scaling: at small $r$, such a self-similarity results from the analytic nature of fields, while at large $r$ it follows from Kolmogorov-like phenomenological arguments.

We now focus more closely on the elasto-inertial range of scales, where the derivation presented in \S\ref{sec:derivation} applies. Fig.~\ref{fig:Fig2}(c) shows that, the ESFs deviate from the expected $\rp/3$ of the solid straight lines for intermediate $r$ (when $\rp \neq 3$). Similar deviations from the K41 predictions are well known to arise from the intermittent nature of turbulence in Newtonian flows, where the structure functions $\Sf{\rp} \sim r^{\xip}$ with $\xip < \rp/3$, for $\rp > 3$~\cite{Frisch96}. In polymeric turbulence, we instead find that $\Stp \sim r^{\zetap}$ with $\zetap > \rp/3$ for $\rp >3$ (notice that the $\Stp$ curves are steeper than the guiding lines at intermediate $r$), hinting that $\widetilde{\delta V}_{\parallel}$ is a relatively less rough signal than the $\delta u_\parallel$ counterpart in classical Newtonian turbulence. This may be understood by recalling that $\widetilde{\delta V}_{\parallel}$ comprises of both differences and averages over a separation $\br$ (see \S\ref{sec:derivation}): since differences emphasise fluctuations, while averages tend to smoothen them, the inertial contribution makes $\widetilde{\dV}_\parallel$ rough, whereas the elastic contribution makes it smoother. This, and a weakening fluid non-linearity with decreasing $r$, implies extended fluctuations are smoother than the velocity fluctuations in Newtonian turbulence, resulting in $\zetap > \xip$.

The deviation of $\zetap$'s from their predicted $\rp/3$ value hints that relation~\eqref{eq:PK41} does not hold exactly for $\rp \neq 3$, and corrections must be accounted for. This can be achieved within the KO62 phenomenology, as discussed in \S\ref{sec:derivation}, by associating the statistics of $\widetilde{\dV}_{\parallel}$ with the scale-averaged total dissipation $\disstr$. To reiterate, the idea is to relax the global averaging of the total dissipation $\bra{\diss_t}$ in Eqn.~\eqref{eq:PK41} in favour of $\bra{\diss_{t,r}}$. This means that, we can simply write the ratio of two different moments $\Stp$ and $\Stq$ of $\widetilde{\delta V}_\parallel$, $\forall \rp, \rq \in \mathbb{R}$, using Eqn.~\eqref{eq:PK61} in the form $\tilde{S}_{\rp} = A_{\rp} \bra{\disstr^{\rp/3}} r^{\rp/3}$, and obtaining
\begin{equation}
\Ht{\rp}{\rq}  \equiv  \frac{\Stilde{\rp} }{\Stilde{\rq}^{\rp/\rq} } \frac{\disstlm{\rq}^{\rp/\rq}}{\disstlm{\rp}}= \frac{A_{\rp}}{A_{\rq}^{\rp/\rq}} = constant.
  \label{eq:Hyper}
\end{equation}
The hyperflatness $\Ht{\rp}{\rq}$, therefore, should be constant in $r$ if the KO62 phenomenology holds for the statistics of the extended observable in polymeric turbulence. For the present discussion, we choose $\rp=6$ and $\rq=2$, and plot $\Ht{6}{2}$ as a function of $r$ in Fig.~\ref{fig:Fig2}(d). Indeed, we find that $\Ht{6}{2} \approx 5$ remains fairly flat in the entire elasto-inertial range of scales. In comparison, hyperflatness $H_{6,2} \equiv \lrp{S_6/S_2^3} \bra{ \dissfr^{2/3 } }^3/ \bra{ \dissfr^{6/3} }$ of the purely Newtonian contribution is not constant, and shows a strong scale dependence, especially in the elastic range of scales. Clearly, the refined Kolmogorov phenomenology works fairly well even in polymeric turbulence, but for the suitably defined extended observable $\widetilde{\delta V}_\parallel$. The curves for the uncompensated ratios $\Stilde{6}/\Stilde{2}^3$ and $S_6/S_2^3$, expectedly, are not flat and are non-trivial functions of $r$. The weaker dependence of $\Stilde{6}/\Stilde{2}^3$ on $r$ than $S_6/S_2^3$ implies that $\widetilde{\dV}_{\parallel}$ is less intermittent than $\delu{\parallel}$, especially in the elastic range. This can again be attributed to a dominant polymeric contribution that tends to make $\widetilde{\dV}_\parallel$ less rough in the elastic range. In the inertial range of scales, where $\fflux$ dominates, the curves for $\lrp{S_6/S_2^3}$ and $\tilde{S}_6/\tilde{S}_2^3$ overlap: the flow elasticity is subdominant here, so that $\widetilde{\delta V}_\parallel \approx \delta u_\parallel$. 

Overall, Fig.~\ref{fig:Fig2}(d) underlines the fact that, the scale averaged total dissipation $\disstr$ determines the statistics of the extended longitudinal velocity increments $\widetilde{\delta V}_\parallel$ in turbulent polymeric flows, as does $\dissfr$ for $\delta u_\parallel$ in classical Newtonian turbulence. In other words, the refined Kolmogorov phenomenology KO62 extends fairly well to polymeric turbulence in the entire elasto-inertial range of scales, provided that the suitable observable $\widetilde{\dV}_\parallel$ is considered. 

After establishing the central role played by $\disstr$ in polymeric turbulence, we now move on to test another of Kolmogorov's hypotheses that intends to free the description of turbulence of the locally averaged dissipation~\citep{K62}.

\subsection{The Kolmogorov multipliers}

\begin{figure}
	\centering
	\includegraphics[width=\textwidth]{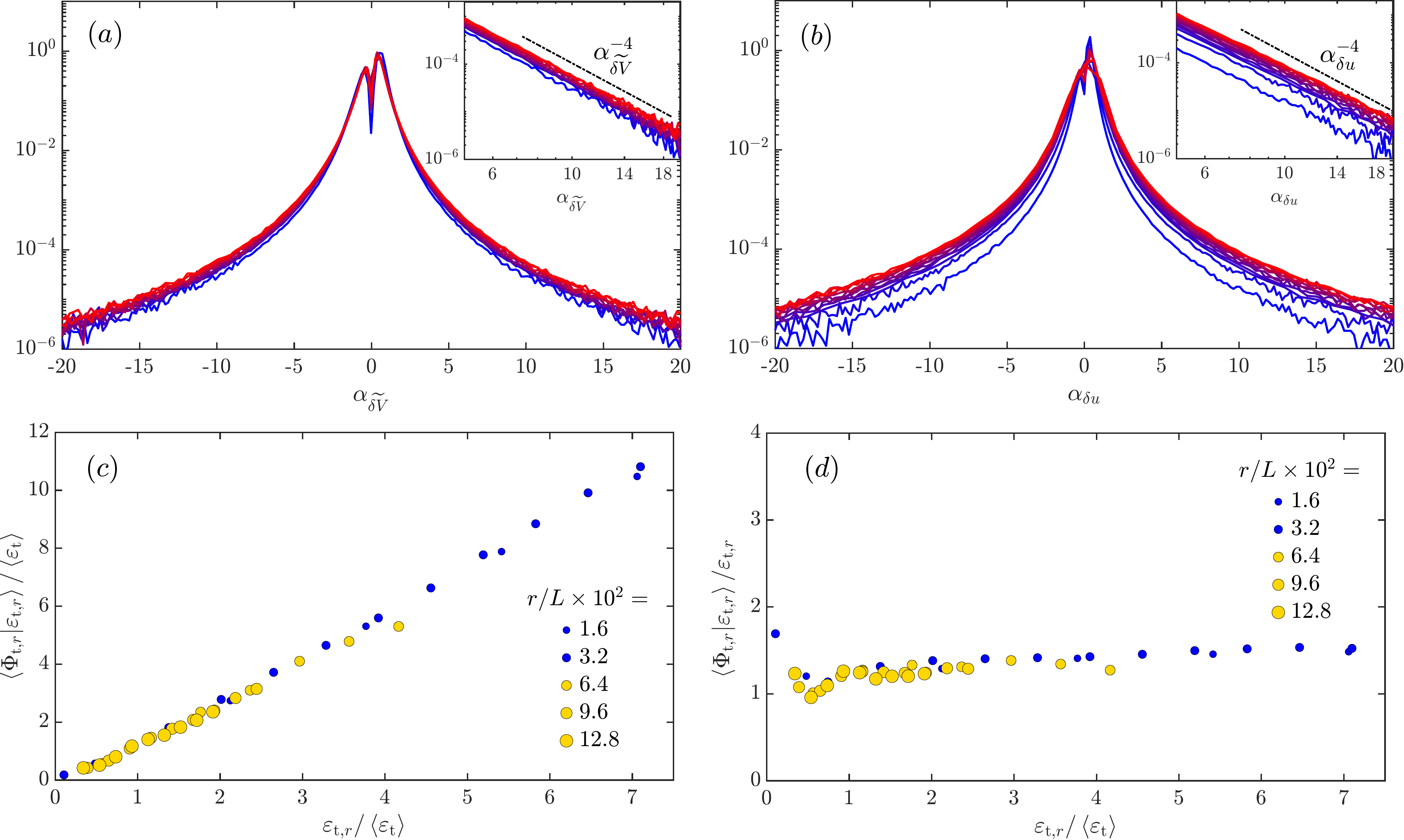} 	
	\caption{\textbf{Kolmogorov Multipliers} \textbf{(a)} Probability distribution functions of the extended Kolmogorov multipliers $\mult{\widetilde{\dV}} $ for scale ratio $r'/r=4$ are scale invariant, and collapse for a wide range of $r$ ($r'$) lying in the elasto-inertial range. \textbf{(b)} The corresponding multipliers for velocity increments $\mult{\delu{}}$ do not collapse. Different colours refer to different values of $r$ $(r' = 4r)$ spanning the entire elasto-inertial range, from the red curve at $r/L \approx 4 \times 10^{-3}$ to the blue curve at $r/L \approx 7.4 \times 10^{-2}$. The insets show the power-law decay of the tails with exponent $-4$, for both extended and Newtonian multipliers. \textbf{Conditionally Averaged Flux} \textbf{(c)} The conditionally averaged total flux $\bra{\Phi_{{\rm t},r} | \disstr} \propto \disstr$ for scales $r$ spanning the elasto-inertial range conforms to the KO62 hypothesis. \textbf{(d)} This means that $\bra{\Phi_{{\rm t},r} | \disstr}/\disstr \sim const.$ in this range.}
	\label{fig:Fig3}
\end{figure}
In his 1962 work, Kolmogorov proposed an alternative similarity hypothesis in an attempt to have a statistical description of turbulence, without resorting to the locally averaged dissipation. The hypothesis states that in the limit $\Rey \to \infty$, the distributions of the Kolmogorov multipliers $\alpha (\bX,r,r')$ defined as the ratio of velocity increments at two different scales $r$ and $r'$, 
\begin{align}
\alpha(\bX,r,r') \equiv \frac{ \delta u_i ( \bm{X}, r \hat{e}_i  ) }{ \delta u_i (\bm{X}, r' \hat{e}_i) },			\label{eq:MultNewt}
\end{align}
are universal functions of only the ratio $r/r'$, and not of the absolute scales themselves~\citep{K62}. For Newtonian turbulence, the distributions of the multipliers have indeed been shown to be scale-invariant~\cite{Sreeni1992,Sreeni1993,Sreeni1995,Sreeni2003}. However, Kolmogorov's additional proposition that the multipliers are log-normally distributed does not hold with the distributions actually exhibiting Lorentzian power-law tails~\citep{Sreeni1992,Sreeni2003}. 

We now show that the distributions of the multipliers based on the extended velocity increments exhibit a scale-invariant behaviour across scales, even in polymeric turbulence. To this end, we introduce the extended $\mult{\widetilde{\dV}}$ as well as the purely Newtonian multipliers as $ \mult{\delu{}}$:
\begin{align}
  \mult{\widetilde{\dV}} \equiv \frac{ \widetilde{\delta V}_\parallel(\bm{X},r) }{\widetilde{\delta V}_\parallel(\bm{X},r')} ;		\quad   
  \mult{\delu{}} \equiv \frac{ \left( \delta u_\parallel (\delta u_j)^2 \right)^{1/3}(\bm{X},r)}{ \left( \delta u_\parallel (\delta u_j)^2 \right)^{1/3}(\bm{X},r')}.
\end{align}
Note that, our Newtonian multipliers differ in their definition from~\eqref{eq:MultNewt}; we make this choice to pick out the purely Newtonian multipliers from the definition~\eqref{eq:dvt} of $\widetilde{\dV}_{\parallel}$. However, we show in the Supplementary Material that our conclusions are independent of the choice of $\mult{\delta u}$ or $\alpha(\bm{X},r,r')$.

Here we fix the ratio $r'/r=4$ and plot the distributions of $\mult{\widetilde{\dV}}$ and $\mult{\delta u}$ in the top panels of Fig.~\ref{fig:Fig3}, for different values of $r$ and $r'$ such that both lie in the elasto-inertial range of scales. We see that the distributions of $\mult{\widetilde{\dV}}$ fall fairly well on top of each other for a wide range of $r$, and show universal power-law tails $\alpha_{\widetilde{\delta V}}^{-4}$. This demonstrates the scale-invariance of $\mult{\widetilde{\dV}}$ in the intermediate range of scales. The same, instead, cannot be said for the Newtonian multipliers $\mult{\delta u}$, whose distributions for different scales do not show a good collapse, and are thus devoid of a scale-invariant nature in the entire elasto-inertial range. 

Overall, with the aid of the extended multipliers, we show that scale-invariance is again restored in polymeric turbulence with respect to the extended observable $\widetilde{\delta V}_\parallel$. 

\subsection{Conditional Fluxes}

In this final section, we provide additional evidence of our arguments by relying on a reformulation of the refined Kolmogorov similarity hypothesis based on the KHMH equation~\citep{Yeung2024}. We start from Eqn.~\eqref{eq:duidui} and average all terms over a sphere of radius $r/2$ in the $\bm{r}-$space at any point $\bm{X}$, to obtain the evolution equation for the local kinetic energy $k_r$ up to scale $r$, defined as $k_r \equiv (1/2\Omega_{r/2}) \int_{\Omega_{r/2}} \text{d}\Omega \ (\delta u_i^2/2) $, where $\Omega_r = (4/3) \pi r^3$. For polymeric turbulence, this equation reads
\begin{align}
\frac{\partial k_r}{\partial t} = \underbrace{\Psi_{{\rm f},r} + \Psi_{{\rm p},r}}_{\Psi_{{\rm t},r}} + \underbrace{ \Phi_{{\rm f},r} + \Phi_{{\rm p},r} }_{ \Phi_{{\rm t},r} } +  
\textrm{P}_r + \textrm{D}_r  + \textrm{F}_r - \underbrace{ \left( \dissfr + \disspr \right) }_{ \disstr},
\end{align}
where $\Psi_{t,r}$ is the total advective rate of change of $k_r$, $\textrm{P}_r$ is the average pressure work, $\textrm{D}_r$ denotes the viscous diffusion of $k_r$, and $\textrm{F}_r$ is the volume averaged energy injection rate. The subscript $r$ refers to locally averaged quantities, which are functions of $\bX,r$ and $t$. We focus on $\Phi_{t,r}$, which is defined as
\begin{align}
  \Phi_{{\rm t},r}(\bm{X},r,t) \equiv \underbrace{- \frac{3}{4 r} \frac{1}{4 \pi r^2} \int_{S(r)} \left( \delta u_i^2 \delta u_j \right) n_j \text{d}S}_{\Phi_{{\rm f},r}(\bm{X},r,t)}  + \underbrace{\frac{3}{4 r} \frac{1}{4 \pi r^2} \int_{S(r)} \left( 4 \delta u_i T_{ij}^* \right) n_j \text{d}S}_{ \Phi_{{\rm p},r}(\bm{X},r,t) },
\end{align}
and is interpreted as the local energy cascade rate at position $\bm{X}$, scale $r$ and time $t$. In this context, the Kolmogorov refined similarity hypothesis implies that in the entire elasto-inertial range of scales the statistics of $\Phi_{t,r}$ depend only on the statistics of $\disstr$. Based on this, the conditional average of the cascade rate $\Phi_{{\rm t},r}$ should obey 
\begin{equation}
  \bra{\Phi_{{\rm t},r}|\disstr} = C_\Phi \disstr,
  \label{eq:RKSH2}
\end{equation}
with $C_\Phi$ being a constant that does not depend on either $r$ or $\disstr$~\citep{Yeung2024}.

%\iffalse
%\begin{figure}
%	\centering
%	\includegraphics[width=0.49\textwidth]{fluxCA_1.eps}
%	\includegraphics[width=0.49\textwidth]{fluxCA_2.eps}
%	\caption{}
%	\label{fig:CondFlux}
%\end{figure}
%\fi
%
In Fig.~\ref{fig:Fig3}(c) and Fig.~\ref{fig:Fig3}(d), we use our numerical data, and plot the measured conditional average $\bra{\Phi_{{\rm t},r}|\disstr}$ as a function of $\disstr$ for different $r$ spanning the entire elasto-inertial range of scales. The analysis considers two scales within the range $1.6 \times 10^{-2} \lesssim r/L \lesssim 3.2 \times 10^{-2}$ that lie in the elastic range, and three scales within $6.4 \times 10^{-2} \lesssim r/L \lesssim 1.3 \times 10^{-1}$ that lie in the inertial range. Evidently, our results show a very good collapse to the expected $\bra{\Phi_{{\rm t},r}|\disstr} = C_\Phi \disstr$ linear behaviour for all $r$, confirming once again that the statistics of the total flux adhere fairly well with the refined Kolmogorov similarity hypothesis, even in polymeric turbulence.

\section{Discussion and Conclusion}

In this work, we present a reconciliation of the turbulence statistics of non-Newtonian, polymeric flows with the Kolmogorov phenomenology. This is achieved by relying on the equivalent K\'{a}rm\'{a}n-Howarth-Monin-Hill (KHMH) relation for turbulent polymeric flows, which encodes the balance of the energy transfer among scales~\citep{deangelis-etal-2005}. The KHMH relation allows us to recognise an invariant range of scales where the net rate of energy transfer, the combination of the non-linear and non-Newtonian fluxes, remains constant. This observation naturally motivates the definition of extended longitudinal velocity increments $\widetilde{\dV}_\parallel$ and the associated structure functions $\Stp \equiv \bra{ \widetilde{\dV}_\parallel^{\rp} }$ allowing to extend the Kolmogorov phenomenology to polymeric turbulence. We show that, $\Stp$ exhibits a self-similar, power-law behaviour in the entire elasto-inertial range of scales as $\Stp \sim r^{\zetap}$, but where $\zetap$ deviates from the predicted $\zetap = \rp/3$ value when $\rp \neq 3$. These deviations, in the spirit of the refined Kolmogorov similarity hypothesis, are well accounted for by considering local scale averages of the total dissipation rather than global averages. Again, in accordance with KO62, the statistics of the total energy flux at a certain scale $r$ depend only on the statistics of the local coarse-grained dissipation $\disstr$. Ultimately, we show that the scale invariance is highlighted by the distributions of the multipliers of $\widetilde{\dV}_\parallel$, which indeed show a good collapse for a wide range of scales. 
 
With the presented formalism, we are able to show that, turbulence statistics of polymeric flows can be cast within the refined Kolmogorov phenomenology after suitable modifications, that naturally arise from the governing equations of motion. Such a formalism is rather general, and may be easily extended to inhomogeneous, anisotropic flows, as well as flows comprising fluid mixtures. 

\bibliography{refs}

\section{Acknowledgments}
The research was supported by the Okinawa Institute of Science and Technology Graduate University (OIST) with subsidy funding to M.E.R. from the Cabinet Office, Government of Japan. M.E.R. also acknowledges funding from the Japan Society for the Promotion of Science (JSPS), grants 24K00810 and 24K17210. The authors acknowledge the computer time provided by the Scientific Computing Section of the Core Facilities at OIST, and the computational resources provided by the HPCI System (Project IDs: hp230018, hp220099, hp210269, and hp210229). %M.E.R. acknowledges useful discussion with Prof. Guido Boffetta, Stefano Musacchio, and Dhrubaditya Mitra.

\section{Author contributions}
A.C. and R.K.S. conceived the original idea. M.E.R. performed the numerical simulations. A.C. and R.K.S. performed the computations and analysed the data. M.E.R. developed the simulation code and supervised the research. All authors outlined the manuscript content and finalised the manuscript, with the first draft written by A.C..

\section{Competing interests}
The authors declare that they have no competing interests.

\section{Data availability}
All data needed to evaluate the conclusions are present in the paper and/or the Supplementary Information.

\section{Code availability}
The code used for the present research is a standard direct numerical simulation solver for the Navier--Stokes equations. Full details of the code used for the numerical simulations are provided in the Methods section and references therein.

\appendix

\pagenumbering{gobble}

\setcounter{table}{0}
\makeatletter 
\renewcommand{\thetable}{S\@arabic\c@table}
\makeatother

\setcounter{figure}{0}
\makeatletter 
\renewcommand{\thefigure}{S\@arabic\c@figure}
\makeatother

\setcounter{equation}{0}
\makeatletter 
\renewcommand{\theequation}{S\@arabic\c@equation}
\makeatother

\section*{Supplementary Discussion}

\section{Polymeric Dissipation}
\label{smat:Pdiss}

In polymeric flows, energy is dissipated by both the carrier fluid and the dissolved polymers. That energy is indeed lost by fluid dissipation is ensured by the positive-definiteness of $\dissf \equiv \nu \lrp{\partial u_i/\partial x_j}^2$, which appears with a negative sign in equation~(4) of the section 2 of the main text. But can the same be said of the polymeric dissipation $\dissp \equiv T_{ij} \partial u_i/\partial x_j$? We show here that, although $\dissp $ admits both positive and negative values locally, its global average $\bra{\dissp}$ remains positive-definite. 

We start our discussion by deriving the equation for the fluid energy $u_i^2/2$, to show that the energy injected by the forcing is dissipated by the combination of the fluid and polymeric dissipations. We multiply equation~(1) on the main text by $u_i$ and, by averaging over the volume using the $\bra{\cdot}$ operator, we obtain
\begin{align}
  \frac{\partial}{\partial t} \bra{ \frac{u_i^2}{2} } +
  \bra{ u_i u_j \frac{\partial u_i}{\partial x_j} } = 
  - \frac{1}{\rho} \bra{ u_i \frac{\partial p}{\partial x_i } } +
  \nu \bra{ u_i \frac{ \partial^2 u_i }{\partial x_j^2 } } +
  \frac{1}{\rho} \bra{ u_i \frac{\partial T_{ij}}{\partial x_j} } +
  \bra{ u_i f_i}.
\end{align}
Here, $\partial \bra{ u_i^2/2 }/\partial t = 0$ due to stationarity, $\bra{u_i f_i} = \bra{\dissi}$ is the energy injection from the forcing, and, taking into account the periodic boundary conditions and the incompressibility constraint, we write
\begin{align}
  \bra{ u_i u_j \frac{\partial u_i}{\partial x_j } } & = \frac{\partial}{\partial x_j} \bra{ u_j \frac{u_i^2}{2} } - \bra{\frac{u_i^2}{2} \frac{\partial u_j}{\partial x_j} } = 0, \nonumber \\
  \frac{1}{\rho}\bra{ u_i \frac{\partial p}{\partial x_i } } & = \frac{1}{\rho} \frac{\partial \bra{ p u_i }}{\partial x_i} - \frac{1}{\rho}\bra{ p \frac{\partial u_i}{\partial x_i} } = 0, \nonumber \\
  \nu \bra{ u_i \frac{\partial^2 u_i}{\partial x_j^2} } & = \frac{\nu}{2} \frac{\partial}{\partial x_j} \bra{ \frac{\partial u_i^2}{\partial x_j} } - \nu \bra{ \frac{\partial u_i}{\partial x_j} \frac{\partial u_i}{\partial x_j} } = - \nu \bra{ \frac{\partial u_i}{\partial x_j} \frac{\partial u_i}{\partial x_j} } = - \bra{ \dissf }, \nonumber \\
  \frac{1}{\rho} \bra{ u_i \frac{\partial T_{ij}}{\partial x_j} } & = \frac{1}{\rho} \frac{\partial}{\partial x_j} \bra{ u_i T_{ij} } - \frac{1}{\rho} \bra{ T_{ij} \frac{\partial u_i}{\partial x_j} } = - \frac{1}{\rho} \bra{ T_{ij} \frac{\partial u_i}{\partial x_j} } = - \bra{\dissp}. \nonumber
\end{align}
Thus, pressure and fluid non-linearity do not contribute to the net energy balance, and the total energy injected by the forcing in the system is balanced by the fluid and polymer dissipations when in a stationary state, i.e. $\bra{\dissi} = \bra{\dissp} + \bra{\dissf}$.

Positive-definiteness of $\bra{\dissp}$ can be argued using equation~(2) on the main text. For simplicity we consider an Oldroyd-B fluid such that $\mathcal{P}=1$. After volume averaging and using the statistically stationarity, the incompressibility constraint, and the periodic boundary conditions, we obtain
\begin{align}
 \bra{ R_{kj} \frac{\partial u_i}{\partial x_k} } + \bra{ R_{ik} \frac{\partial u_k}{\partial x_j} } = \frac{1}{\tau_p} \bra{ R_{ij} - \delta_{ij} }.
 \label{eq:pol2}
\end{align}
Next, we take the trace of the previous equation
\begin{align}
  \bra{ R_{ki} \frac{\partial u_i}{\partial x_k} } + \bra{ R_{ik} \frac{\partial u_k}{\partial x_i} } = 2 \bra{ R_{ij} \frac{\partial u_i}{\partial x_j} }=  \frac{1}{\tau_p} \bra{ R_{jj} - 3},
\end{align}
and we use the definition $R_{ij} = T_{ij} \tau_p/\mu_p + \delta_{ij}$ to eventually obtain
\begin{align}
 \bra{2 \frac{\tau_p}{\mu_p} T_{ij} \frac{\partial u_i}{\partial x_j} } 
 + \underbrace{ 2 \bra{ \delta_{ij} \frac{\partial u_i}{\partial x_j} } }_{ = 0} = 
 \frac{1}{\tau_p} \bra{ R_{jj} - 3 } = \frac{1}{\mu_p} \bra{ T_{jj} }.
 \label{eq:pol3}
\end{align}
Equation~\eqref{eq:pol3} implies that $\bra{ \dissp } \equiv \bra{ T_{ij} \partial u_i/\partial x_j } = \lrp{\mu_{\rp}/2 \taup^2}  \bra{ R_{jj} -3 } = \bra{ T_{jj} }/2 \tau_p$~\cite{MER2023}. Since $R_{jj}-3$ and $T_{jj}$ are positive-definite quantities, $\bra{\dissp} \ge 0$; see also e.g.~\cite{deangelis-etal-2005}.

\section{The numerical database}

To support our theoretical arguments, we rely on a database of homogeneous, isotropic, polymeric turbulence obtained with direct numerical simulations, that has been introduced by~\citep{RKS2023}. Their paper contains full details on the numerical method and on the computational procedures, which are only briefly recalled here. Equations~(1) and (2) of the main text are numerically integrated using the in-house solver \href{https://groups.oist.jp/cffu/code}{\textit{Fujin}}, which uses an incremental pressure-correction scheme. The Navier--Stokes equations written in primitive variables are solved on a staggered grid using second-order finite-differences in all the directions. The momentum equation is advanced in time using a second-order Adams-Bashforth time scheme, while the polymer conformation tensor $R_{ij}$ is advanced with a second-order Crank-Nicolson scheme. A log-conformation formulation~\citep{Marco18} ensures positive-definiteness of the conformation tensor at all times. Turbulence is sustained using the Arnold-Beltrami-Childress (ABC) cellular-flow forcing \citep{podvigina-pouquet-1994}. The equations are solved within a cubic domain of size $L=2\pi$ with periodic boundary conditions in all directions, discretised with $1024$ grid points in each direction, to ensure that all the scales down to the smallest dissipative ones are properly solved, i.e. $\eta/\Delta x = \mathcal{O}(1)$, where $\Delta x$ is the grid spacing and $\eta$ is the Kolmogorov scale. The parameters are chosen to achieve in the purely Newtonian case a microscale Reynolds number of $Re_\lambda = \urms \lambda/\nu \approx 460$. The fluid and polymer dynamic viscosities are fixed such that $\mu_f/(\mu_f+\mu_p) = 0.9$. Simulations are advanced with a constant time step of $\Delta t/\tau_\eta = 2 \times 10^{-3}$, where $\tau_\eta$ is the Kolmogorov time scale.

\section{The Third Order Extended Structure Function $\Stilde{3}$}

\begin{figure}[!ht]
	\centering
	\includegraphics[width=0.5\textwidth]{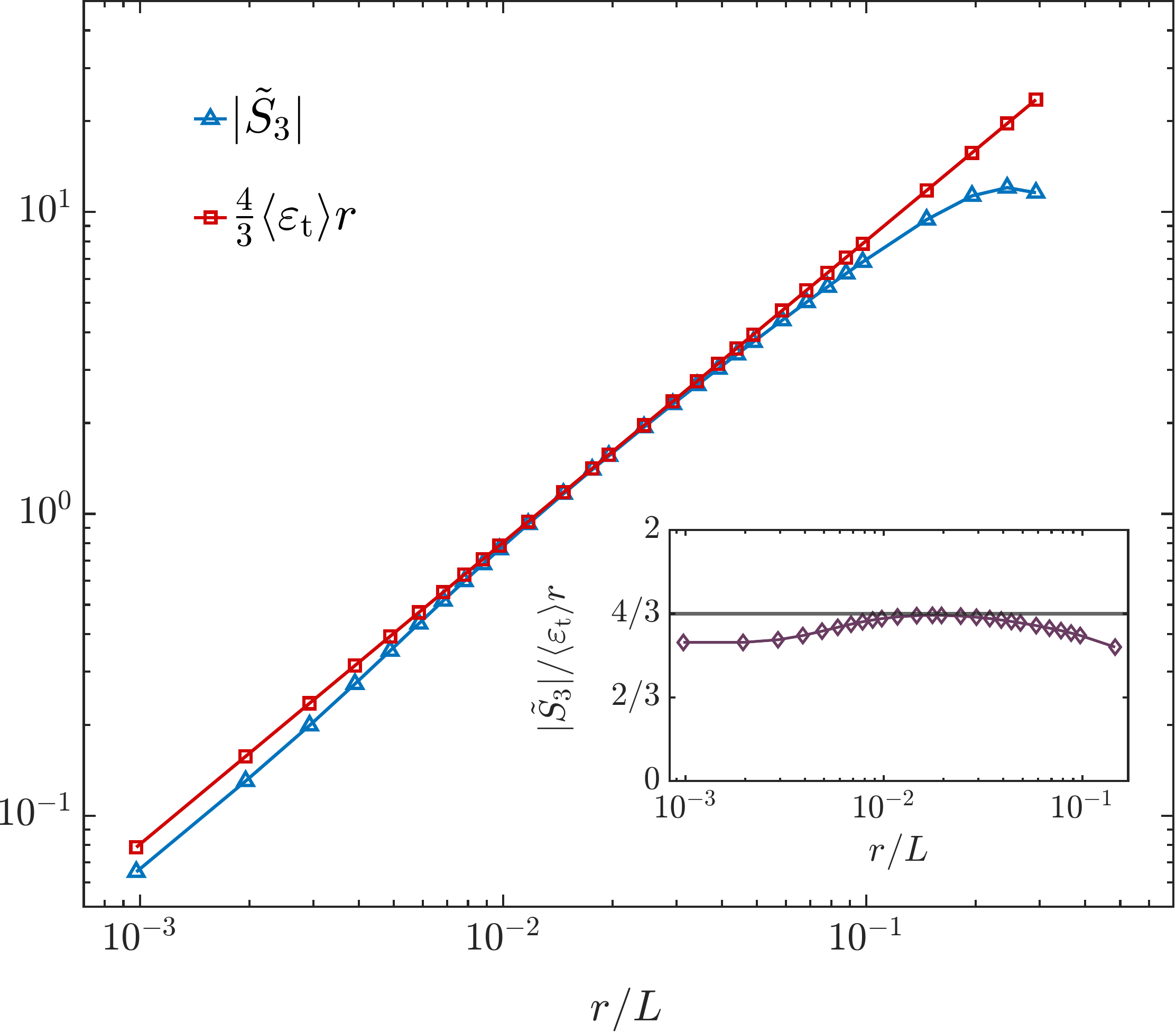}
	\caption{\textbf{(main)} The (absolute of) $\Stilde{3}$ compared to $\lrp{4/3} \bra{\disst} r$. The curves fall on top of each-other in an intermediate range of scales, justifying the validity of the extended Kolmogorov theory to polymeric turbulence. \textbf{(inset)} The ratio $\lrv{\Stilde{3}}/\bra{\disst}r$ equals 4/3 in the same range, as a further confirmation of our results. } 
	\label{fig:CompS3t}
\end{figure}

In this section we comment further on the exactness of equation~(9) in the main text. We recall that relation (9) states that:
\begin{align}
	\Stilde{3}(r) = -\frac{4}{3} \bra{\disst} r 		\quad \implies   \quad 		\frac{\lrv{\Stilde{3}(r)}}{\bra{\disst}r}  = \frac{4}{3}.	\label{eq:CompS3t}
\end{align}
We show in the main panel of Fig.~\ref{fig:CompS3t} the plots corresponding to both the right and the left hand sides of~\eqref{eq:CompS3t}. Both the curves fall on top of each-other in an intermediate range of scales, where our Kolmogorov-like relation~\eqref{eq:CompS3t} is expected to hold. We also plot in the inset the ratio $\lrv{\Stilde{3}}/\bra{\disst}r$, which must equal 4/3 in the intermediate range of scales. The ratio is indeed equal to 4/3 (shown as a solid black line) for almost half a decade of scales. 

\section{Analyticity in the dissipation range}
\label{smat:Analyticity}

We have shown in the main text how the KHMH relation yields a Kolmogorov $\rp/3$ scaling for the extended structure functions $\Stp = \bra{\dV_{\parallel}^{\rp}}$ in the elasto-inertial range of scales of polymeric turbulence, where the dissipative and forcing terms are negligible (see section \Romannum{3} A). Curiously enough, we observe the same self-similar power-law behaviour also at the smallest scales, where the dissipative term can not be neglected. This can be seen in Fig.2(a) of the main text where the $\Stp$ curves become parallel to the straight lines as $r \to 0$. We now show below that such a behaviour results from the analyticity of the fields at small scales that indeed gives rise to a $\rp/3$ exponent.

At small enough scales, all the dynamical fields must be analytic functions of the coordinates. That is, they can be Taylor expanded as a power series in separation $r$. A simple consequence is that the scaling exponents of the classical structure functions are trivial, i.e. $\sp \sim r^{\rp}$~\cite{Schumacher07}. Of course, this is also true in our non-Newtonian flows where $S_2 \sim r^2$ for $r \rightarrow 0$ as shown in Fig.1(a) of the main text. We therefore simply Taylor expand our velocity and polymer fields assuming analyticity (indices are suppressed for clarity):
\begin{align}
	u(x+r) &= u(x) + r \frac{\partial u}{\partial x} + \mathcal{O}(r^2)  = u(x) + A r + \hot	\quad     \text{and} \nonumber \\
	T(x+r) &= T(x) + r \frac{\partial T}{\partial x} + \mathcal{O}(r^2)  = T(x) + B' r + \hot,	\nonumber 
\end{align}
where $\hot$ stands for higher order terms. We can therefore write the velocity increments $\delta u$ and the averages of the polymeric extra-stress tensor $T^*$ as:
\begin{align}
	\delta u = \diffu = Ar + \hot \ ,\qquad \text{and} \qquad T^* = \avgT = T(x) + Br + \hot,
\end{align}
where $B=B'/2$. Using these relations at the leading order in the definition of $\langle \widetilde{\dV}^3 \rangle$, we obtain to leading order:
\begin{align}
	\bra{\widetilde{\dV}^3_\parallel} = \bra{A^3 r^3 - 4 A r (T+Br)} + \hot =  \bra{A r \lrs{( - 4 (T + Br ) - A^2 r^2} }  + \hot,
\end{align}
which implies that
\begin{align}
	\bra{\widetilde{\dV}^3_\parallel}  \sim r^{1} 	\quad 	 \text{for} 	\quad	 r \rightarrow 0;
\end{align}
where in the last step we have neglected $\mathcal{O}(r^2)$ terms and higher. One can obtain in a similar vein that $\Stp \sim \bra{\widetilde{\dV}^{\rp}_\parallel} \sim r^{\rp/3}$. The $\rp/3$ exponent at small $r$ is thus a consequence of the analyticity of the fluid velocity and the polymer fields. 
% This explain the $\Stp \equiv \langle \widetilde{\delta V}^{\rp} \rangle \sim r^{\rp/3}$ scaling at small $r$ observed in figure 2 of the main text.

\section{Kolmogorov multipliers}
\label{smat:KM}

\begin{figure}[!ht]
	\centering
	\includegraphics[width=\textwidth]{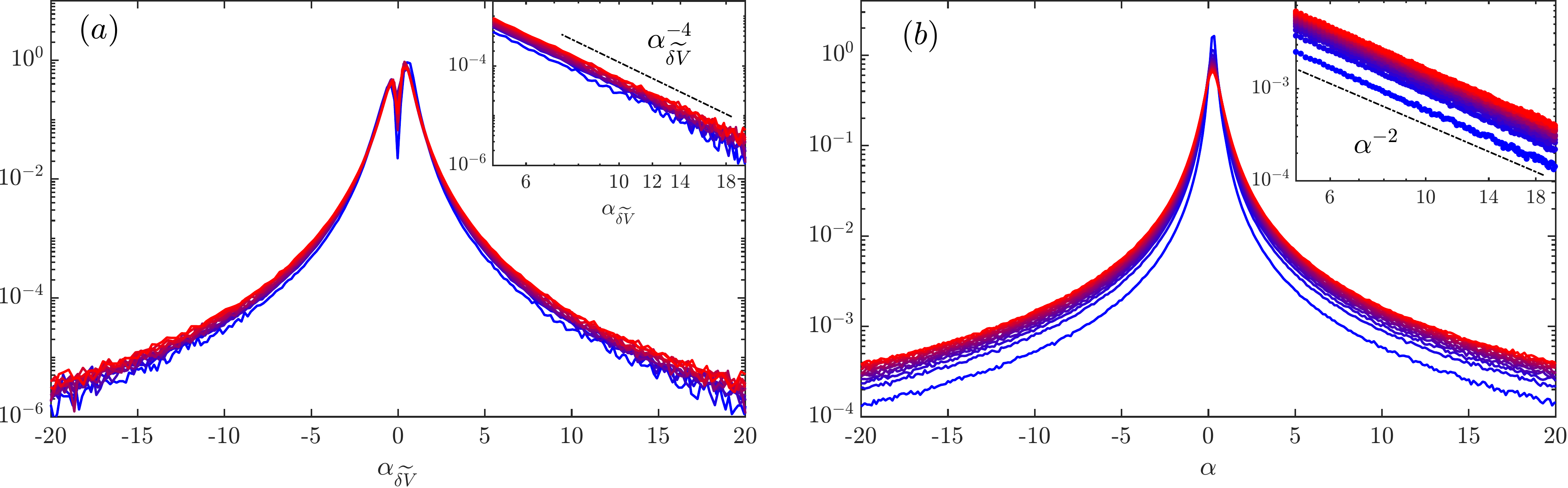}
	\caption{\textbf{Kolmogorov multipliers with scale ratio $\bm{r/r'=4}$.} Moving from blue to red colour, the $r$ and $r'$ values increase, from $r/L \approx 1.6 \times 10^{-2}$ to $r/L \approx 2.0 \times 10^{-1}$. The first three (blue) lines have both points in the elastic range of scales where $\pflux > \fflux$ (see Fig.~1 of the main text). \textbf{(a)} Multiplier distributions for extended velocity increments $\alpha_{\widetilde{\delta V}}$ reproduced from the main text. The black dashed line in the inset sows a power-law decay of tails as $\alpha_{\widetilde{\delta V}}^{-4}$. \textbf{(b)} Distribution of the classical multipliers $\alpha$. The dash-dotted line in the inset show a power-law fall off of tails with exponent ${-2}$.} 
	\label{fig:Mult}
\end{figure}

In the main text, we have used the extended Kolmogorov multipliers to demonstrate the self-similar behaviour of $\widetilde{\delta V}_\parallel$ in the elasto-inertial range of scales. This entailed comparing the extended and the Newtonian multipliers, that were defined as:
\begin{align}
  \mult{\widetilde{\dV}} \equiv \frac{ \widetilde{\delta V}_\parallel(\bm{X},r) }{\widetilde{\delta V}_\parallel(\bm{X},r')} ;		\qquad   
  \mult{\delu{}} \equiv \frac{ \left( \delta u_\parallel (\delta u_j)^2 \right)^{1/3}(\bm{X},r)}{ \left( \delta u_\parallel (\delta u_j)^2 \right)^{1/3}(\bm{X},r')}.
  \label{Mult}
\end{align}
It was shown that, unlike $\mult{\delta u}$, the distributions of $\mult{\widetilde{\dV}}$ collapse fairly well for a wide range of scales, provided both $r$ and $r'$ fall in the elasto-inertial range of scales. Such a definition was inspired by the fact that, $\alpha_{\delta u} = \alpha_{\widetilde{\delta V}}$ in the absence of the polymers . 

We now show that, the very same conclusions follow from a more classical definition of the Newtonian multipliers,
\begin{equation}
\mult{} \equiv \frac{ \delta u_i( \bX,r e_i ) }{ \delta u_i( \bX,r'e_i ) },				\label{ClassMult}
\end{equation}
that was used in the past to demonstrate the near scale invariance of multiplier statistics in purely Newtonian turbulence~\citep{K62,Sreeni2003}. We plot the probability distribution functions (pdfs) of $\alpha(\bX,r,r')$ computed using this definition in Fig.~\ref{fig:Mult}(b). The distributions, yet again, do not collapse for different values of $r$ (and $r'$), unlike those of $\mult{\widetilde{\delta V}}$ which fall on top of each other for the same $r,r'$.  Therefore, we again have that scale-invariance in polymeric turbulence, over the complete elasto-inertial range of scales, can be expected only for the extended fluctuation $\widetilde{\delta V}_{\parallel}$. This modified definition means that the $\alpha$-distributions now have Lorentzian $\alpha^{-2}$ tails, similar to what was already observed for classical Newtonian turbulence~\cite{Sreeni2003}. Note, however, that definitions in~\eqref{Mult} yield power-law tails that scale as $\alpha^{-4}$ (see also section \Romannum{3} B in the main text).

\section{Model and parameter dependence}

\begin{figure}[!ht]
  \centering
  \includegraphics[width=\textwidth]{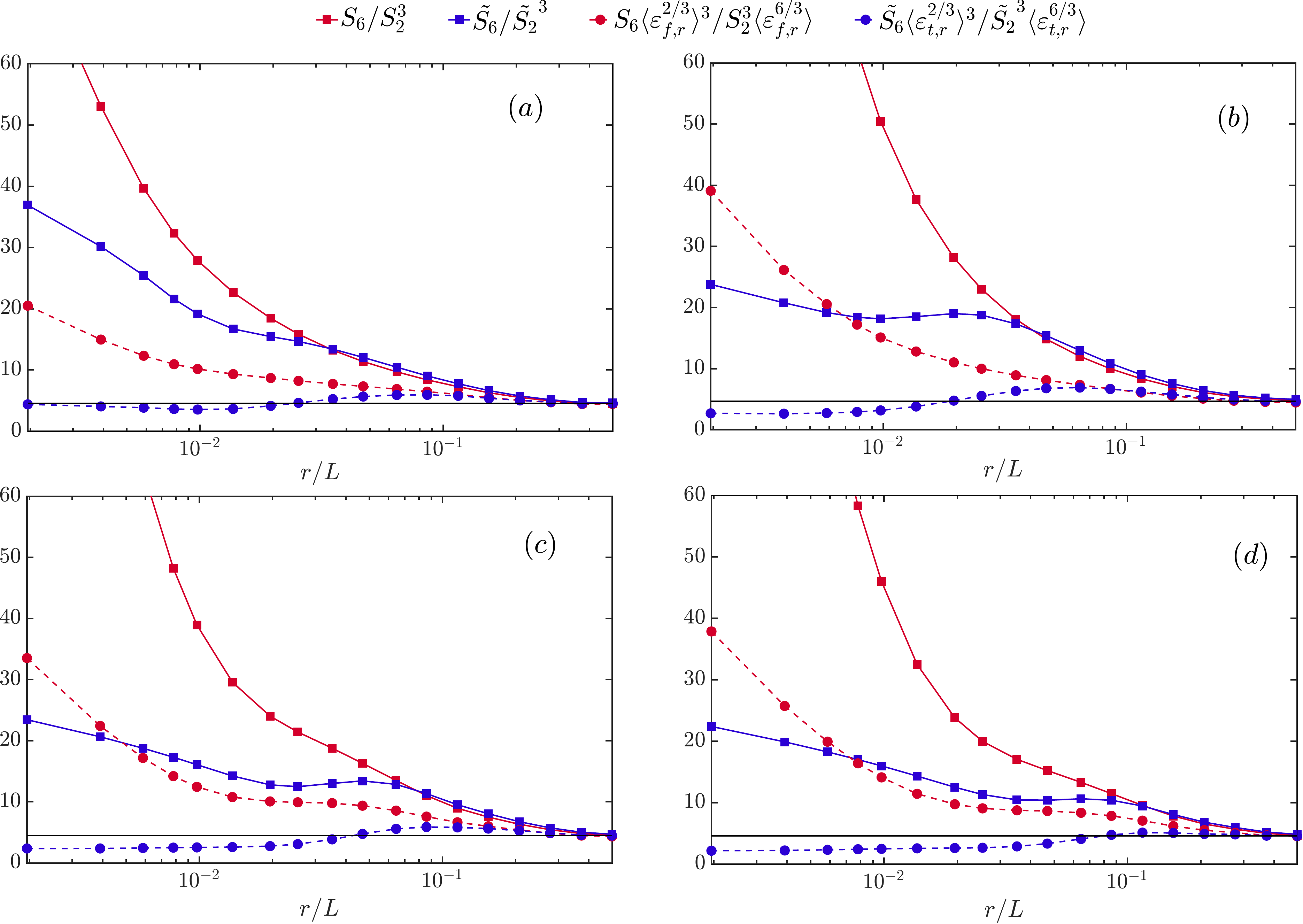}
  \caption{\textbf{Dependence of the compensated extended structure function on model and elasticity}. \textbf{Top:} The dependence on the polymer model. \textbf{(a)} Structure functions for the FENE-P model with $\Deb=1$. \textbf{(b)} Structure functions from for the Oldroyd-B simulations at De = 1. \textbf{Bottom:} Dependence on polymer elasticity. Oldroyd-B structure functions for \textbf{(c)} $\Deb=3$ and \textbf{(d)} $\Deb=9$ .}
  \label{fig:De_Mod_dep}
\end{figure}    

In the main text, we consolidate our analytical arguments via data from simulations of turbulent, Oldroyd-B polymeric fluids. We now show here that, our results and conclusions are independent of the polymeric model and are a consequence of the fluid elasticity. We perform additional direct numerical simulations at the same $\Rel \approx 460$ with the polymers now modelled as Finitely Extensible Non-linear Elastic (FENE) springs (with the Peterlin approximation), FENE-P. We compare the results obtained from the Oldroyd-B simulations with those from the FENE-P at the same $\Deb=1$ in the top panel of Fig.~\ref{fig:De_Mod_dep}. The left panel corresponds to FENE-P polymers and the right to Oldroyd - B. The $\Stp$ for FENE-P simulations are more intermittent than those from Oldroyd-B in the dissipation range, but less so in the elasto-inertial range. This is seen from the plots of uncompensated $\Stp$s shown with blue square markers. These curves, upon compensation with suitable powers of $\disstr$, become almost flat in both the cases (see the lines with blue circles), confirming that our results do not depend on the polymeric model.

We also establish here that our conclusions presented for $\Deb = 1$ in the main text are not particular to just one instance of polymer elasticity, but remain valid across $\Deb$ numbers. We illustrate this in the bottom panels of Fig.~\ref{fig:De_Mod_dep}, where we plot the (uncompensated and compensated) curves of Newtonian and extended structure functions for different polymer elasticity $\Deb = 3$ (left) and $\Deb = 9$ (right) obtained from Oldroyd-B simulations. Fig.~\ref{fig:De_Mod_dep} shows that, the $\tilde{S}_6/\tilde{S}_2^3$ curve collapses fairly well to a constant value in the entire range of $r$ for all cases, when compensated with the local total dissipation, confirming thus the validity of our results. As expected, the $\tilde{S}_6/\tilde{S}_2^3$ curves start deviating from the classical $S_6/S_2^3$ ones at larger $r$, when increasing $\Deb$, consistently with the presence of a wider elastic range of scales where $\pflux > \fflux$. 

\section{Decomposing the polymeric contribution}
\begin{figure}[!ht]
	\centering
	\includegraphics[width=0.54\textwidth]{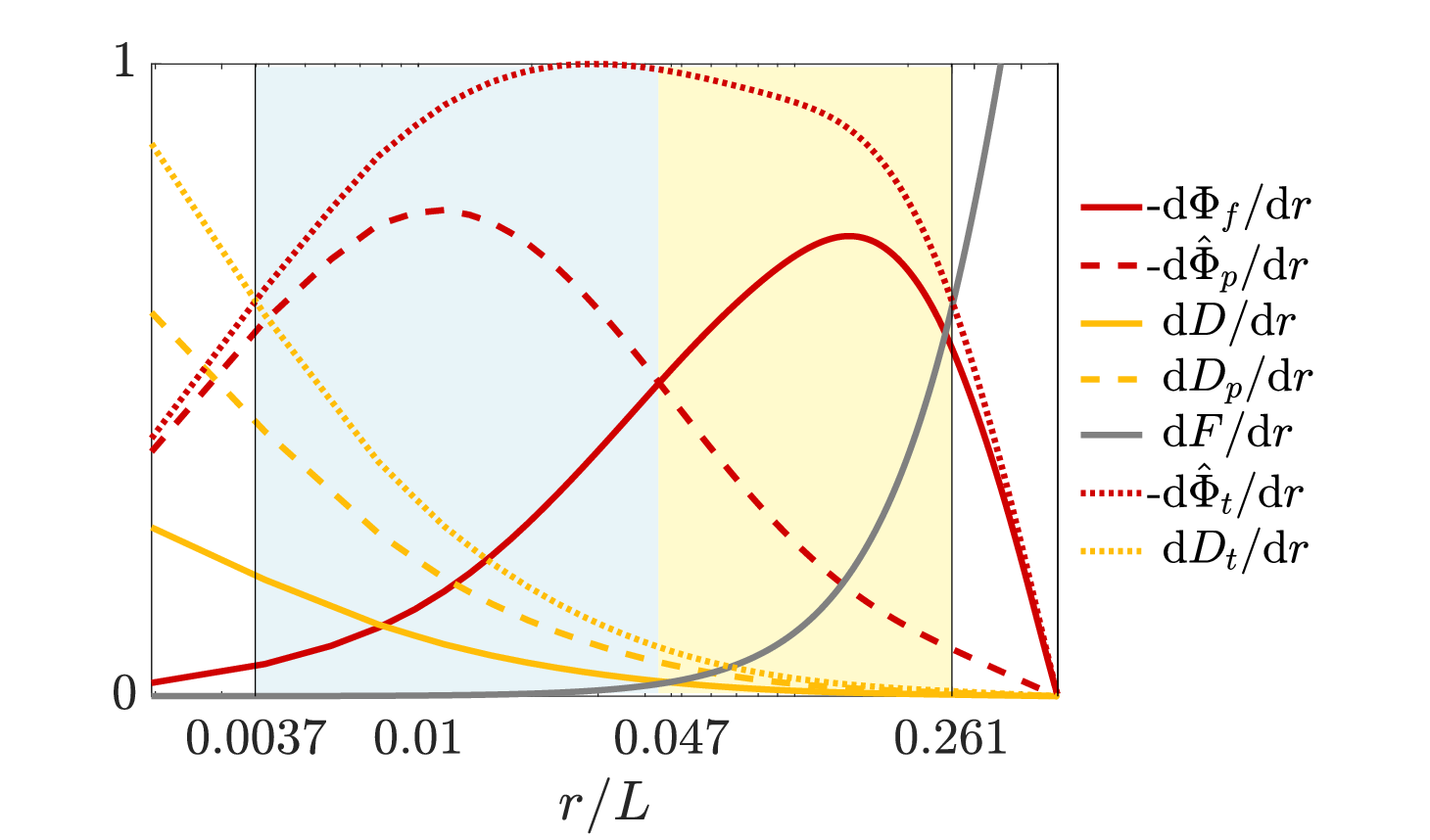}
	\caption{\textbf{Decomposing the polymeric contribution.} The normalised flux contributions to the KHMH, as a function of scale $r$, when the dissipative polymeric flux $\rD_{\rp}$ is removed from the polymeric flux $\pflux$ using~\eqref{fig:phipdec}. We show in blue (yellow) the range of scales dominated by $\hat{\Phi}_{\rp}$ ($\fflux$). The dashed red line represents the total flux $\hat{\Phi}_{\rm t}$.}
	\label{fig:budg_mod}
\end{figure}
Polymers provides an additional mode of dissipation of energy in turbulent flows which was denoted by $\dissp$ in the main text. This polymeric dissipation remains hidden in the non-Newtonian flux $\pflux$ and manifests itself as only at the smallest scales. We now separate the purely dissipative contribution from $\pflux$ using the same heuristic arguments used in~\cite{MER2023}. This entails modelling the polymer dissipative flux contribution $\rD_{\rp}$ in a qualitatively similar fashion to fluid dissipation flux $\rD$. This contribution $\rD_{\rp}$ does not contribute to the transfer of energy across scales. We assume that, for $r \rightarrow 0$, $\rD_{\rp}$ has the same asymptotic dependence as $\rD(r)$, and that in this limit $\text{d}\rD_{\rp}/\text{d}r = (4/3)\bra{\dissp} $. We can thus write the simple relation:
\begin{align}
  \pflux(r) = \hat{\Phi}_{\rp}(r) + \rD_{\rp}(r),
  \label{fig:phipdec}
\end{align}
where 
\begin{align}
 \rD_{\rp}(r) = \frac{ \bra{\dissp} }{ \bra{\dissf} } \rD(r).
\end{align}

The remaining total flux $\text{d}\hat{\Phi}_{\rm t}/\text{d}r = \text{d}(\hat{\Phi}_{\rp} + \Phi_{\rm f})/\text{d}r$ exhibits a clear plateau in the elasto-inertial range of scales in Fig.~\ref{fig:budg_mod}. Expectedly, the forcing ($F$) and the dissipative terms ($\rD$ and $\rD_{\rp}$) remain negligible in this range. $\hat{\Phi}_{\rp}$ and $\fflux$ intersect at $r/L \approx 0.047$, with $\fflux$ being dominant at larger scales and $\hat{\Phi}_{\rp}$ at smaller ones. We use the crossover scale $r/L \approx 3.6 \times 10^{-3}$ between $\hat{\Phi}_{\rm t}$ and $\rD_{\rm t} = \rD + \rD_{\rp}$ to define the dissipative range of scales, and the intersection between $\hat{\Phi}_{\rm t}$ and $F$ at $r/L \approx 0.26$ to delimit the large scales where energy is injected.
\begin{figure}
	\centering
	\includegraphics[width=0.49\textwidth]{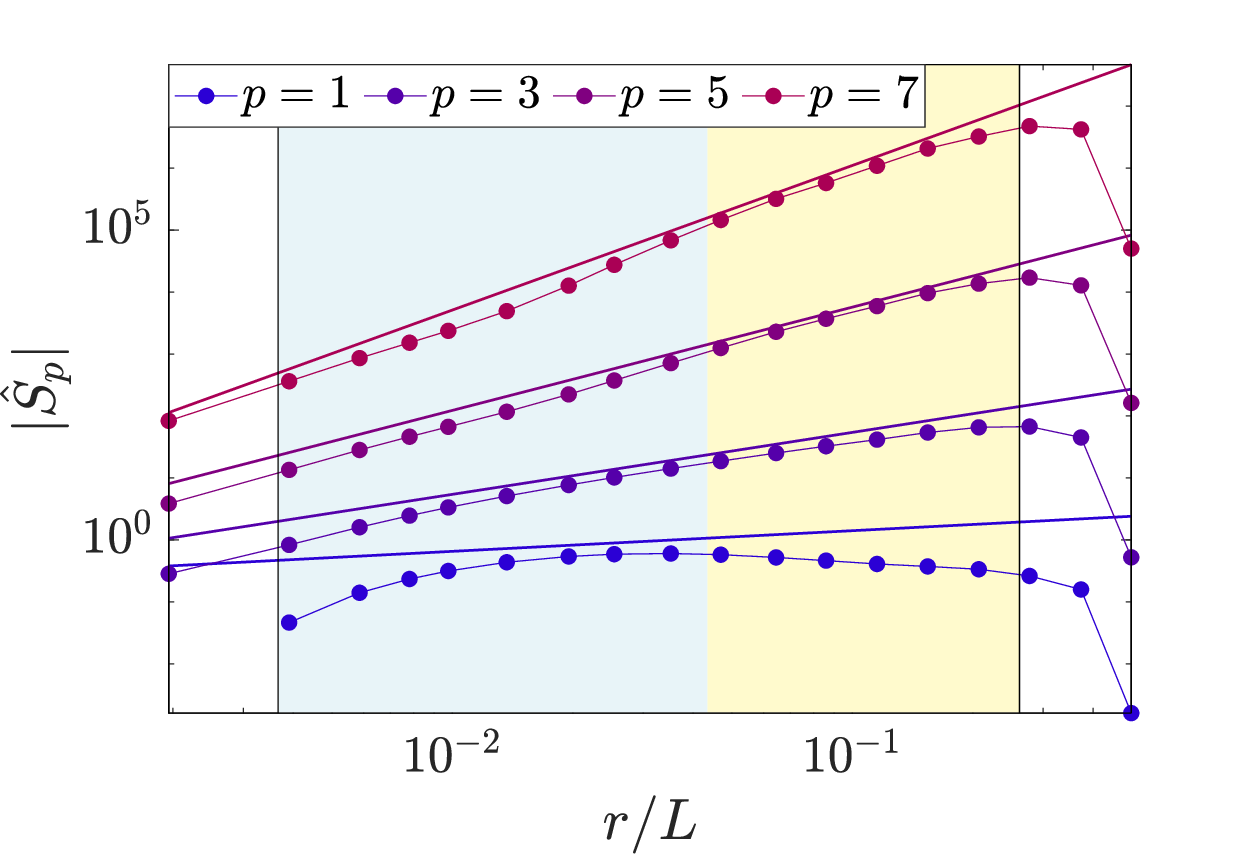}
	\includegraphics[width=0.49\textwidth]{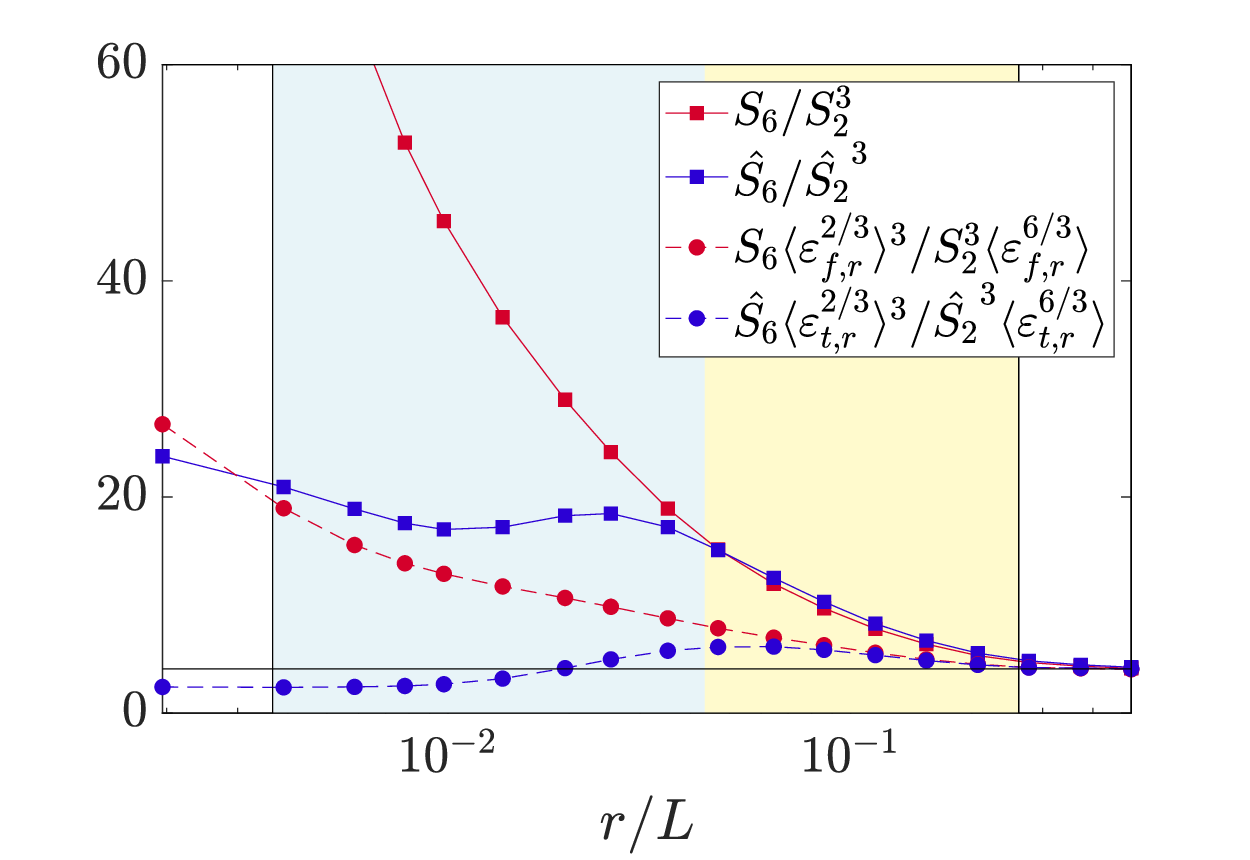}
	\caption{\textbf{Modified extended structure functions.} The log-log plots of the modified, extended structure functions $\hat{S}_{\rp}$ vs the scale $r$, for various orders $\rp$. Same as Fig.~2 of the main text, but for $\widehat{\delta V}_{\parallel}$. \textbf{(left)} Odd-order extended structure functions based on $\widehat{\delta V}_{\parallel}$. \textbf{(right)} Compensated $\widehat{\delta V}_{\parallel}$ extended structure functions.}
	\label{fig:comp_mod}
\end{figure}
Using the decomposition~\eqref{fig:phipdec} in equation~(6) of the main text, the budget in the complete elasto-inertial range of scales is modified to
\begin{align}
  \hat{\Phi}_{\rm t} \equiv \fflux(r) + \hat{\Phi}_{\rp}(r) = - \frac{4}{3} \bra{ \diss_{\rm t}} r;
\end{align} 
here, we have ignored both fluid $\rD$ and polymeric dissipative $\rD_{\rp}$ contribution to the flux balance. By using local isotropy, we then obtain
\begin{align}
  \bra{ \delta u_\parallel (\delta u_i)^2 } - 4 \bra{ \delta u_i T_{i\parallel}^* } +
  2 \nu \frac{\bra{\dissp}}{\bra{\dissf}} \frac{\partial \bra{(\delta u_i)^2 }}{\partial r_\parallel}  = 
  - \frac{4}{3} \bra{ \disst } r.
\end{align}
This relation facilitates an alternate definition of the extended $\rp{th}-$order structure function $\hat{S}_{\rp} \equiv \bra{\widehat{\delta V}_\parallel^{\rp}} $, where the alternative extended velocity fluctuation $\widehat{\delta V}_\parallel$ is defined as:
\begin{align}
  \widehat{\delta V}_\parallel = \left( \delta u_\parallel (\delta u_i)^2 - 4 \delta u_i T_{i\parallel}^* 
                              + 2 \nu \frac{\bra{\dissp}}{\bra{\dissf}} \frac{ \partial (\delta u_i)^2}{ \partial r_\parallel}  \right)^{1/3}.
\end{align}

Fig.~\ref{fig:comp_mod} clearly shows that all our arguments hold also when considering $\widehat{\delta V}_{\parallel}$ instead of $\widetilde{\delta V}_{\parallel}$, with the compensated $\hat{S}_6/\hat{S}^3_2$ collapsing fairly well to a constant value in the entire inertial range of scales.

\section{Distribution of the extended velocity increments}

\begin{figure}
\centering
\includegraphics[width=\textwidth]{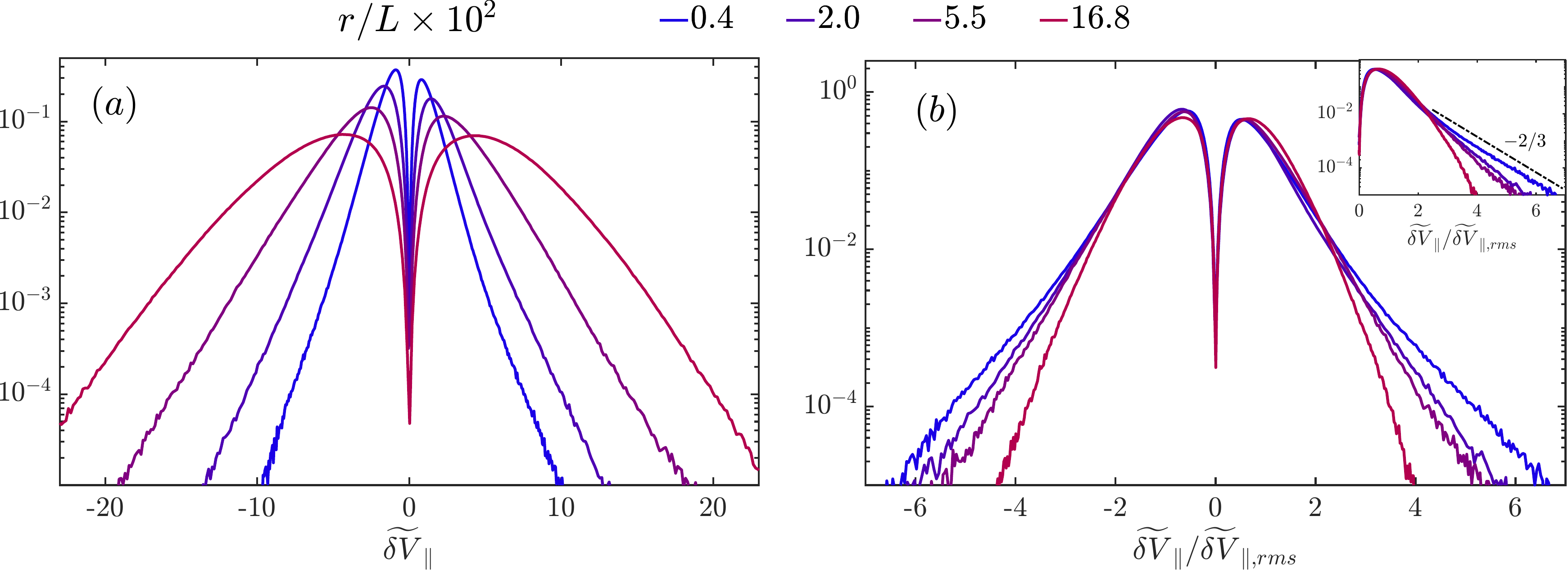}
\caption{\textbf{Pdfs of the extended velocity increment $\widetilde{\delta V}_\parallel$ at different scales $r$}. \textbf{(a)} The pdfs show wider spread at large scales, as the differences grow large for large separations. \textbf{(b)} The normalized fluctuations show intermittent behaviour, with wider tails at smallest scales. The inset shows the exponential decay of the tails.}
\label{fig:pdfdvt}
\end{figure}

It is straightforward to see from the definition (12) in the main text that the extended increments $\widetilde{\dV}_\parallel$ differ significantly in their characteristics from the Newtonian velocity increments. This is indeed seen in Fig.~\ref{fig:pdfdvt} which shows the distributions of $\widetilde{\delta V}_\parallel(r)$ for different scales $r$ that range from the elastic scales where $\pflux$ dominates ($3.6 \times 10^{-3} \lesssim r/L \lesssim 4.7 \times 10^{-2}$), to the inertial scales where $\fflux$ dominates ($4.7 \times 10^{-2} \lessapprox r/L \lessapprox 2.6 \times 10^{-1}$). Panel (a) shows bare $\widetilde{\dV}$ distributions while panel (b) shows the distributions for the normalized (by the root mean square) variable. The probability of $\widetilde{\delta V}_\parallel \approx 0$ is severely suppressed at large $r$ due to the fact that even though $\delta u$ is small, the exact condition with $\delta u_\parallel (\delta u_i)^2 = 4 \delta u_i T_{i\parallel}$ is much less likely to occur. In panel (b), the distributions are right skewed for small $r$, with the mode being negative and with the right tail being slightly more pronounced. When increasing $r$ the distribution becomes progressively more symmetric. Also, the relevance of the tails changes with $r$ (see the right panel). For small $r$ the tails are more relevant, and the $\widetilde{\delta V}_\parallel$ signal is more intermittent. For large scales, instead, the tails decay much faster and the signal is much less intermittent. The symmetry of the distribution for large scales explains the dependence of $\tilde{S}_1$ on $r$, with the positive/negative cancellation being the cause of the depletion of $\tilde{S}_1$ at large scales (see Fig. 2(a) in the main text). We also show in the inset of panel (b) that the distributions at all scales show an exponential decay so that their moments or $\Stilde{\rp}$ are all well-defined.

\end{document}